\documentclass[11pt]{article}
\usepackage{amsfonts,amsmath}

\def\bbt{\bibitem}
\def\be{\begin{equation}}
\def\en{\end{equation}}
\def\ber{\begin{eqnarray}}
\def\enr{\end{eqnarray}}
\def\nmb{ \nonumber\\}
\def\d{\partial}

\def\ov{\over }
\def\tld{\tilde}

\def\sgm{\sigma}
\def\Sgm{\Sigma}
\def\al{\alpha}
\def\bet{\beta}
\def\gm{\gamma}
\def\Gm{\Gamma}
\def\im{\imath}

\def\lm{\lambda}
\def\Lm{\Lambda}
\def\Om{\Omega}

\def\et{\eta}
\def\tt{\theta}

\def\ups{\upsilon}
\def\dlt{\delta}
\def\Dl{\Delta}
\def\kp{\kappa}

\def\bh{{\bf h}}
\def\bp{{\bf p}}

\def\bt{{\bf t}}

\def\bv{{\bf v}}
\def\bd{{\bf d}}
\def\bw{{\bf w}}
\def\bs{{\bf s}}
\def\be{{\bf e}}
\def\bmu{{\boldsymbol \mu}}
\def\bLm{{\boldsymbol \Lm}}

\def\blm{{\boldsymbol \lambda}}
\def\bphi{{\boldsymbol \phi}}

\begin{document}
%\nopagenumbers
%\rightline{Landau Tmp.}
\vskip 2 true cm
\centerline{\bf FREE-FIELD APPROACH TO D-BRANES}
\centerline{\bf IN GEPNER MODELS.}
\vskip 1.5 true cm
\centerline{\bf S. E. Parkhomenko}
\centerline{Landau Institute for Theoretical Physics}
\centerline{142432 Chernogolovka, Russia}
\vskip 0.5 true cm
\centerline{spark@itp.ac.ru}
\vskip 1 true cm
\centerline{\bf Abstract}
\vskip 0.5 true cm

 We represent free-field construction of boundary states in Gepner models
basing on the free-field realization of N=2 superconformal
minimal models. Using this construction we consider the open string spectrum
between the boundary states
and show that it can be described in terms of Malikov, Schechtman, Vaintrob
chiral de Rham complex of the Landau-Ginzburg orbifold.
It allows to establish direct relation of the open string spectrum
for boundary states in Gepner models to the open string spectrum
for fractional branes in Landau-Ginzburg orbifolds. The example of $1^{3}$ model
considered in details.
\vskip 10pt

"{\it PACS: 11.25Hf; 11.25 Pm.}"


\smallskip
\vskip 10pt
\centerline{\bf0. Introduction}
\vskip 10pt

 Dirichlet branes give realization of solitonic states in string theory which have
Ramond-Ramond
charges and play important role in non-perturbative aspects of the theory ~\cite{Pol1}.
The description of $D$-branes on curvered string backgrounds in Calabi-Yau (CY) models 
of superstring
compactification is one of the important problems of string theory.

 At large volume the $D$ branes can be described by classical geometric techniques of
bundles on submanifolds of
Calabi-Yau manifold. The extrapolation into the stringy regime usually requires boundary
conformal field
theory (BCFT) methods. In this approach $D$-brane configurations are
given by conformally invariant boundary states or boundary conditions.
The boundary states in Gepner models
have been considered first by Recknagel and Schomerus ~\cite{ReS}.

 In contrast to the large volume limit the quantum string generalization of the corresponding
geometric objects is developed in much less extent
and most of the investigations devoted to the boundary state approach to $D$-branes are
related to the problem of the string generalization of the geometry. The considerable
progress in the understanding of the quantum geometry of $D$-branes at small volume of
the CY manifold has been achieved mainly due to the works ~\cite{ReS}-~\cite{Dou1}.
The main idea developed in these papers is to relate the intersection index of boundary states
with the bilinear form of the $K$-theory classes of bundles on the large volume CY
manifold and use this relation to associate the $K$-theory classes to the boundary states
establishing thereby the correspondence between the boundary states and bundles on
CY manifold. The most important results of further studies has been the
uncovering in the approach of linear sigma model ~\cite{LSM}
of the special role of exceptional sheaves on CY manifolds and their relation to the
boundary states in Gepner models ~\cite{DD}, ~\cite{May}-~\cite{Tom}.
The extension of this picture to the arbitrary points in the moduli space
has been mainly developed in ~\cite{Dou}, ~\cite{AsLo}, ~\cite{Shar}.

 In this paper we consider another aspect of boundary states in Gepner models coming from
$N=2$ minimal models construction ~\cite{Gep},~\cite{EOTY}
and represent free-field approach to their construction and study.
Our goal is to try to establish by the free-fields direct relation of the geometric 
properties of $D$-branes with vertex algebra structures of BCFT description.

 Being building blocks of Gepner models, $N=2$ superconformal minimal models 
represent a subclass of
rational CFT where the construction of the boundary states
leaving a whole chiral symmetry algebra unbroken can
be given in principle and the interaction of these states
with closed strings can be calculated exactly. But in practice the calculation of closed string
amplitudes in general CFT backgrounds is available only if the corresponding free-field realization of
the model is known. Therefore, it is important to extend free-field approach to 
the case of rational models of CFT with a
boundary. This problem has been treated recently in  ~\cite{SP}-~\cite{Kaw},
where free-field realizations of degenerate $sl(2)$ Kac-Moody and $N=2$ Virasoro algebra
representations  ~\cite{BFeld}-~\cite{FeS} has been used to explicit boundary states construction 
and boundary correlation function calculation.

 We extend in this paper the free-field construction ~\cite{SP1} of boundary states in $N=2$
minimal models to the case of boundary states in Gepner models and show that free-field
representation appears to be very useful for investigation of geometry of boundary states.

 In section 1 we review free-field construction of irreducible representations in
$N=2$ minimal models developed by Feigin and Semikhatov. In section 2 the free-field
realization of Gepner models is briefly discussed. Section 3 is devoted to the explicit
free-field construction of Ishibashi states. In section 4 free-field representations of
$A$ and $B$-type Recknagel Shomerus boundary states in Gepner models are given and the
open string spectrum of states is calculated using free-field realization of Ishibashi states.
In section 5 we investigate the open string spectrum between the boundary states
using the ideas of vertex operator algebra approach to the string theory on toric CY
manifolds recently developed in ~\cite{MSV}-~\cite{B1}.
The detailed consideration of the open string spectrum is carried out in the simplest example of
$1^{3}$ Gepner model which appears to be quite representative to illustrate the idea how 
to establish by the free-fields the direct relation of the Gepner models boundary states to
the fractional branes in Landau-Ginzburg orbifolds and toric geometry of CY manifolds.
More detailed investigation
of the geometry of boundary states and the relation with the results of ~\cite{ReS}-~\cite{Dou},
~\cite{May}-~\cite{Tom}
we hope to develop in future publication.

\vskip 10pt
\centerline{\bf 1. Free-field realization of $N=2$ minimal models}
\centerline{\bf irreducible representations.}
\vskip 10pt

 In this section we briefly discuss free-field construction of
Feigin and Semikhatov ~\cite{FeS} of the irreducible modules
in $N=2$ superconformal minimal models. Free-field approach to $N=2$ minimal models
considered also in \cite{Ohta}-~\cite{Wit}.

\leftline{\bf 1.1. Free-field representations of $N=2$ super-Virasoro
algebra.}

  We introduce (in the left-moving sector) the free bosonic fields
$X(z), X^{*}(z)$ and free fermionic fields $\psi(z), \psi^{*}(z)$,
so that its OPE's are given by
\ber
X^{*}(z_{1})X(z_{2})=\ln(z_{12})+reg.,\nmb
\psi^{*}(z_{1})\psi(z_{2})=z_{12}^{-1}+reg,
\label{1.ope}
\enr
where $z_{12}=z_{1}-z_{2}$. Then for an arbitrary number
$\mu$ the currents of $N=2$ super-Virasoro
algebra are given by
\ber
G^{+}(z)=\psi^{*}(z)\d X(z) -{1\ov \mu} \d \psi^{*}(z), \
G^{-}(z)=\psi(z) \d X^{*}(z)-\d \psi(z), \nmb
J(z)=\psi^{*}(z)\psi(z)+{1\ov \mu}\d X^{*}(z)-\d X(z), \nmb
T(z)=\d X(z)\d X^{*}(z)+
{1\ov 2}(\d \psi^{*}(z)\psi(z)-\psi^{*}(z)\d \psi(z))-\nmb
{1\ov 2}(\d^{2} X(z)+{1\ov \mu}\d^{2} X^{*}(z)),
\label{1.min}
\enr
and the central charge is
\ber
c=3(1-{2\ov \mu}).
\label{1.cent}
\enr

 As usual, the fermions in NS sector are expanded into
half-integer modes:
\ber
\psi(z)=\sum_{r\in 1/2+Z}\psi[r]z^{-{1\ov 2}-r},\
\psi^{*}(z)=\sum_{r\in 1/2+Z}\psi^{*}[r]z^{-{1\ov 2}-r},\nmb
G^{\pm}(z)=\sum_{r\in 1/2+Z}G^{\pm}[r]z^{-{3\ov 2}-r},
\label{1.NS}
\enr
and they are expanded into integer modes in R sector:
\ber
\psi(z)=\sum_{r\in Z}\psi[r]z^{-{1\ov 2}-r},\
\psi^{*}(z)=\sum_{r\in Z}\psi^{*}[r]z^{-{1\ov 2}-r},\nmb
G^{\pm}(z)=\sum_{r\in Z}G^{\pm}[r]z^{-{3\ov 2}-r}.
\label{1.R}
\enr
The bosons $X(z),X^{*}(z),J(z),T(z)$ are expanded in both sectors into integer
modes:
\ber
\d X(z)=\sum_{n\in Z}X[n]z^{-1-n},\
\d X^{*}(z)=\sum_{n\in Z}X^{*}[n]z^{-1-n},\nmb
J(z)=\sum_{n\in Z}J[n]z^{-1-n},\
T(z)=\sum_{n\in Z}L[n]z^{-2-n}.
\label{1.Bex}
\enr

 In NS sector $N=2$ Virasoro superalgebra is acting naturally in Fock module
$F_{p,p^{*}}$ generated by the fermionic
operators $\psi^{*}[r]$, $\psi[r]$, $r<{1\ov2}$, and bosonic operators
$X^{*}[n]$, $X[n]$, $n<0$ from the vacuum state $|p,p^{*}>$ such that
\ber
\psi[r]|p,p^{*}>=\psi^{*}[r]|p,p^{*}>=0, r\geq {1\ov 2},\nmb
X[n]|p,p^{*}>=X^{*}[n]|p,p^{*}>=0, n\geq 1, \nmb
X[0]|p,p^{*}>=p|p,p^{*}>, \
X^{*}[0]|p,p^{*}>=p^{*}|p,p^{*}>.
\label{1.vac}
\enr
It is a primary state with respect
to the $N=2$ Virasoro algebra
\ber
G^{\pm}[r]|p,p^{*}>=0, r>0, \nmb
J[n]|p,p^{*}>=L[n]|p,p^{*}>=0, n>0, \nmb
J[0]|p,p^{*}>={j\ov \mu}|p,p^{*}>=0, \nmb
L[0]|p,p^{*}>={h(h+2)-j^{2}\ov 4\mu}|p,p^{*}>=0,
\label{1.hwv}
\enr
where $j=p^{*}-\mu p$, $h=p^{*}+\mu p$. The vacuum state
$|p,p^{*}>$ corresponds to the vertex operator
$V_{(p,p^{*})}(z)\equiv\exp(pX^{*}(z)+p^{*}X(z))$ placed at $z=0$.

 It is easy to
calculate the character $f_{p,p^{*}}(q,u)$ of the Fock module
$F_{p,p^{*}}$. By the definition
\ber
f_{p,p^{*}}(q,u)=Tr_{F_{p,p^{*}}}(q^{L[0]-{c\ov 24}}u^{J[0]}).
\label{1.fchd}
\enr
Thus we obtain
\ber
f_{p,p^{*}}(q,u)=
q^{{h(h+2)-j^{2}\ov 4\mu}-{c\ov24}}u^{{j\ov \mu}}
{\Theta(q,u)\ov \eta(q)^{3}},
\label{1.fchc}
\enr
where we have used the Jacoby theta-function
\ber
\Theta(q,u)=
q^{1\ov 8}\sum_{m\in Z}q^{{1\ov2}m^{2}}u^{-m}
\label{1.Jthet}
\enr
and the Dedekind eta-function
\ber
\eta(q)=q^{1\ov 24}\prod_{m=1}(1-q^{m}).
\label{1.eta}
\enr

 The $N=2$ Virasoro algebra has the following set of automorphisms
which is known as spectral flow ~\cite{SS}
\ber
G^{\pm}[r]\rightarrow G_{t}^{\pm}[r]\equiv G^{\pm}[r\pm t], \nmb
L[n]\rightarrow L_{t}[n]\equiv L[n]+t J[n]+t^{2}{c\ov 6}\dlt_{n,0},\
J[n]\rightarrow J_{t}[n]\equiv J[n]+t {c\ov 3}\dlt_{n,0},
\label{1.flow}
\enr
where $t\in Z$. Note that spectral flow is intrinsic property
of $N=2$ super-Virasoro algebra and hence, it does not depend on a
particular realization. Allowing in (\ref{1.flow}) $t$ to be
half-integer, we obtain the isomorphism between the NS and R
sectors.

 The spectral flow action on the free
fields can be easily described if we bosonize fermions $\psi^{*}, \psi$
\ber
\psi(z)=\exp(-y(z)), \ \psi^{*}(z)=\exp(+y(z)).
\label{1.fbos}
\enr
and introduce spectral flow vertex operator
\ber
U^{t}(z)=\exp(-t(y+{1\ov \mu}X^{*}-X)(z)).
\label{1.vflow}
\enr
The following OPE's
\ber
\psi(z_{1})U^{t}(z_{2})=
z_{12}^{t}:\psi(z_{1})U^{t}(z_{2}):, \
\psi^{*}(z_{1})U^{t}(z_{2})=
z_{12}^{-t}:\psi^{*}(z_{1})U^{t}(z_{2}):, \nmb
\d X^{*}(z_{1})U^{t}(z_{2})=z_{12}^{-1}t U^{t}(z_{2})+r.,\
\d X(z_{1})U^{t}(z_{2})=-z_{12}^{-1}{t \ov \mu}U^{t}(z_{2})+r.
\label{1.2}
\enr
give the action of spectral flow on the modes of the free-fields
\ber
\psi[r]\rightarrow \psi[r-t], \
\psi^{*}[r]\rightarrow \psi^{*}[r+t], \nmb
X^{*}[n]\rightarrow X^{*}[n]+t\dlt_{n,0}, \
X[n]\rightarrow X[n]-{t\ov \mu}\dlt_{n,0}.
\label{1.3}
\enr

 The action of the spectral flow on the vertex operator
$V_{(p,p^{*})}(z)$ is given by the normal ordered product of the
vertex $U^{t}(z)$ and $V_{p,p^{*}}(z)$. It follows from (\ref{1.3}) that
spectral flow generates twisted sectors.

\leftline{\bf 1.2. Irreducible $N=2$ super-Virasoro representations
and butterfly resolution.}

 The $N=2$ minimal models
are characterized by the condition that
$\mu$ is integer and $\mu\geq 2$. In NS sector the irreducible
highest-weight modules, constituting the (left-moving) space of states of
the minimal model, are unitary and labeled by two integers $h,j$,
where $h=0,...,\mu-2$ and $j=-h,-h+2,...,h$.
The highest-weight vector $|h,j>$ of the module satisfies the conditions
(which are similar to (\ref{1.hwv}))
\ber
G^{\pm}[r]|h,j>=0, r>0, \nmb
J[n]|h,j>=L[n]|h,j>=0, n>0, \nmb
J[0]|h,j>={j\ov \mu}|h,j>, \nmb
L[0]|h,j>={h(h+2)-j^{2}\ov 4\mu}|h,j>.
\label{1.hw}
\enr
If in addition to the conditions (\ref{1.hw}) the relation
\ber
G^{+}[-1/2]|h,j>=0
\label{1.chw}
\enr
is satisfied we call the vector $|h,j>$ and the module $M_{h,j}$
chiral highest-weight vector (chiral primary state) and chiral module,
correspondingly. In this case we have $h=j$.
Analogously, anti-chiral highest-weight vector (anti-chiral primary
state) and anti-chiral
module can be defined if instead of (\ref{1.chw})
\ber
G^{-}[-1/2]|h,j>=0
\label{1.achw}
\enr

 The Fock modules are highly reducible representations of $N=2$ Virasoro algebra.
To see this we introduce following to ~\cite{FeS} two fermionic screening currents
$S^{\pm}(z)$ and the screening charges $Q^{\pm}$
\ber
S^{+}(z)=\psi^{*}\exp(X^{*})(z), \
S^{-}(z)=\psi\exp(\mu X)(z), \nmb
Q^{\pm}=\oint dz S^{\pm}(z)
\label{1.chrg}
\enr
The screening charges commute with the generators of $N=2$ super-Virasoro algebra
(\ref{1.min}). Moreover they are
nilpotent and mutually anticommute
\ber
(Q^{+})^{2}=(Q^{-})^{2}=\{Q^{+},Q^{-}\}=0.
\label{1.BRST}
\enr
But they do not act within each Fock module. Instead they relate to each other the 
different Fock modules. The space where the screening charges are acting can be constructed 
as follows. One has to
introduce the two-dimensional lattice of the momentums:
\ber
\pi=\{(p,p^{*})|p={n\ov\mu},p^{*}=m, n,m\in Z\}
\label{1.L}
\enr
and associate to this lattice the space
\ber
F_{\pi}=\oplus_{(p,p^{*})\in \pi}F_{p,p^{*}}.
\label{1.fock}
\enr

 Due to the properties (\ref{1.BRST}) one can combine the charges $Q^{\pm}$ into
BRST operator acting in $F_{\pi}$ and build a BRST complex
consisting of Fock modules $F_{p,p^{*}}\in F_{\pi}$ such that its cohomology
is given by NS sector of $N=2$ minimal model irreducible module $M_{h,j}$.
This complex has been constructed in ~\cite{FeS}.

 Let us consider first free-field construction
for the chiral module $M_{h,h}$. In this case the complex
(which is known due to Feigin and Semikhatov as butterfly resolution)
can be represented by the following diagram
\ber
\begin{array}{ccccccccccc}
&&\vdots &\vdots &&&&&&\\
&&\uparrow &\uparrow &&&&&&\\
\ldots &\leftarrow &F_{1,h+\mu} &\leftarrow
F_{0,h+\mu}&&&&&&\\
&&\uparrow &\uparrow &&&&&&\\
\ldots &\leftarrow &F_{1,h} &\leftarrow F_{0,h}&&&&&&\\
&&&&\nwarrow&&&&&\\
&&&&&F_{-1,h-\mu}&\leftarrow &F_{-2,h-\mu}&\leftarrow&\ldots\\
&&&&&\uparrow &&\uparrow&\\
&&&&&F_{-1,h-2\mu}&\leftarrow &F_{-2,h-2\mu}&\leftarrow&\ldots\\
&&&&&\uparrow &&\uparrow &&\\
&&&&&\vdots &&\vdots &&
\end{array} \nmb
\label{1.but}
\enr
We shall denote this resolution by $C_{h}$ and denote by $\Gm$
the set where the momentums of the Fock spaces of the resolution take
values.
The horizontal arrows in this diagram are given by the action of
$Q^{+}$ and vertical arrows are given by the action of $Q^{-}$.
The diagonal arrow at the middle of butterfly resolution
is given by the action of $Q^{+}Q^{-}$ (which equals $-Q^{-}Q^{+}$
due to (\ref{1.BRST})). Ghost number operator $g$ of the complex
is defined for an arbitrary vector $|v_{n,m}>\in
F_{n,m\mu+h}$ by
\ber
g|v_{n,m}>=(n+m)|v_{n,m}>, \ if \ n,m\geq 0, \nmb
g|v_{n,m}>=(n+m+1)|v_{n,m}>, \ if \ n,m< 0.
\label{1.grad}
\enr
For an arbitrary vector of the complex $|v_{N}>$ with the
ghost number $N$ the differential $d_{N}$ is defined
by
\ber
d_{N}|v_{N}>=(Q^{+}+Q^{-})|v_{N}>, \ if \ N\neq -1, \nmb
d_{N}|v_{N}>=Q^{+}Q^{-}|v_{N}>, \ if \ N=-1.
\label{1.diff}
\enr
and rises the ghost number by 1. Note that when $N=-1$ one has to
take $d_{N}=Q^{-}Q^{+}=-Q^{+}Q^{-}$ which does not affect the
cohomology.

The main statement of ~\cite{FeS} is that the complex (\ref{1.but}) is exact
except at the $F_{0,h}$ module, where the cohomology is given by
the chiral module $M_{h,j=h}$.

 The butterfly resolution allows to write the character
$\chi_{h}(q,u)\equiv Tr_{M_{h,h}}(q^{L[0]-{c\ov 24}}u^{J[0]})$ of the
module $M_{h,h}$ as the Euler characteristic of the complex:
\ber
\chi_{h}(q,u)=\chi^{(l)}_{h}(q,u)-\chi^{(r)}_{h}(q,u), \nmb
\chi^{(l)}_{h}(q,u)=\sum_{n,m\geq 0}(-1)^{n+m}f_{n,h+m\mu}(q,u),\nmb
\chi^{(r)}_{h}(q,u)=\sum_{n,m> 0}(-1)^{n+m}f_{-n,h-m\mu}(q,u),
\label{1.char}
\enr
where $\chi^{(l)}_{h}(q,u)$ and $\chi^{(r)}_{h}(q,u)$ are the characters
of the left and right wings of the resolution.

 To get the resolutions for other (anti-chiral and non-chiral) modules
one can use the observation ~\cite{FeS} that all irreducible modules can be obtained
from the chiral modules $M_{h,j=h}$, $h=0,...,\mu-2$ by the spectral flow
action $U^{-t}, t=h,h-1,...1$. Equivalently one can restrict the set of chiral
modules by the range $h=0,...,[{\mu\ov2}]-1$ and extend the spectral flow
action by
$t=\mu-1,...,1$ (when $\mu$ is even and $h=[{\mu\ov2}]-1$ the spectral flow
orbit becomes shorter: $t=[{\mu\ov2}]-1,...,1$)~\cite{FeSST}. Thus
the set of irreducible modules can be labeled also by the set
$\{(h,t)|h=0,...,[{\mu\ov2}]-1,\ t=\mu-1,...,0 \}$,
except the case when $\mu$ is even and the spectral flow orbit becomes
shorter.
It turns out that one can get
all the resolutions by the spectral flow action also.
Indeed, the charges
$Q^{\pm}$ commute with spectral flow operator $U^{t}$ as it is
easy to see from (\ref{1.vflow}) and the corresponding OPE's,
moreover, $U^{t}$ is not BRST-exact ($U(z)$ corresponds to the
anti-chiral primary field)
hence, the resolutions
in NS sector are generated from (\ref{1.but}) by the operators
$U^{-t}$, where $t=h,h-1,...,1$ or $t=\mu-1,...,1$
(or $t=[{\mu\ov2}]-1,...,1$ if $\mu$ is even and $h=[{\mu\ov2}]-1$).

 Due to this discussion it is more convenient to change the notation for irreducible
modules. In what follows we shall denote the irreducible modules as $M_{h,t}$,
indicating by $t$ spectral flow parameter.

 As well as the modules and resolutions one can get the characters by the
spectral flow action ~\cite{FeS}:
\ber
\chi_{h,t}(q,u)=q^{{c\ov6}t^{2}}u^{{c\ov3}t}
\chi_{h}(q,uq^{t}).
\label{1.tchi}
\enr
There are the following important automorphism properties of irreducible modules and characters
~\cite{FeS},~\cite{FeSST}.
\ber
M_{h,t}\equiv M_{\mu-h-2,t-h-1}, \
\chi_{h,t}(q,u)=\chi_{\mu-h-2,t-h-1}(q,u),
\label{1.reflect}
\enr
\ber
M_{h,t}\equiv M_{h,t+\mu}, \ \chi_{h,t+\mu}(q,u)=\chi_{h,t}(q,u),
\label{1.oddper}
\enr
where $\mu$ is odd and
\ber
M_{h,t}\equiv M_{h,t+\mu}, \ \chi_{h,t+\mu}(q,u)=\chi_{h,t}(q,u),\ h\neq [{\mu\ov2}]-1, \nmb
M_{h,t}\equiv M_{h,t+[{\mu\ov2}]}, \
\chi_{h,t+[{\mu\ov2}]}(q,u)=\chi_{h,t}(q,u),\ h=[{\mu\ov2}]-1,
\label{1.evper}
\enr
where $\mu$ is even.

  Note that the butterfly resolution
is not periodic under the spectral flow as opposed to the
characters. It is also not invariant with respect to the automorphism (\ref{1.reflect}).
Instead, the periodicity and invariance are recovered on the level of
cohomology. Thus, $U^{\pm\mu}$ spectral flow and automorphism (\ref{1.reflect}) are the
quasi-isomorphisms of complexes.

 The modules, resolutions and characters in R sector are
generated from the modules, resolutions and characters in NS sector by the
spectral flow operator $U^{-{1\ov2}}$.

\vskip 10pt
\centerline{\bf 2. Free-field realization of Gepner model.}
\vskip 10pt

\leftline {\bf 2.1.Free-field realization of the product of
minimal models.}

 It is easy to generalize the free-field representation of the Sect.1. to the case of
tensor product of $r$ $N=2$ minimal models which can be characterized
by $r$ dimensional vector $\bmu=(\mu_{1},...,\mu_{r})$,
where $\mu_{i}\geq 2$ and integer.

 Let $E$ be a real $r$ dimensional vector space and let $E^{*}$
be the dual space to $E$. Let us denote by $<,>$
the natural scalar product in the direct sum $E \oplus E^{*}$.
In the subspaces $E$ and $E^{*}$ we fix the sets of basic vectors
$\Re$ and $\Re^{*}$
\ber
\Re=\{{\bf s}_{i}, i=1,...,r\},
\nmb
\Re^{*}=\{\mu_{i}{\bf s}^{*}_{i}, i=1,...,r\},
\nmb
<{\bf s}_{i},{\bf s}^{*}_{j}>=\dlt_{i,j}.
\label{2.bases}
\enr

 According to the $\Re$ and $\Re^{*}$
we introduce (in the left-moving sector)
the free bosonic fields $X_{i}(z), X^{*}_{i}(z)$ and free
fermionic fields $\psi_{i}(z), \psi^{*}_{i}(z)$, $i=1,...,r$
so that its singular OPE's are given by (\ref{1.ope}) as well as
the following fermionic screening currents and their charges
\ber
S^{+}_{i}(z)={\bf s}_{i}\psi^{*}\exp({\bf s}_{i}X^{*})(z), \nmb
S^{-}_{i}(z)={\bf s}^{*}_{i}\psi\exp(\mu_{i}{\bf s}^{*}_{i}X)(z), \nmb
Q^{\pm}_{i}=\oint dz S^{\pm}_{i}(z).
\label{2.screen}
\enr
For each $i=1,...,r$ one can define by the formulas (\ref{1.min})
N=2 $c_{i}=3(1-{2\ov \mu_{i}})$
Virasoro superalgebra
\ber
G^{+}_{i}={\bf s}_{i}\psi^{*} {\bf s}^{*}_{i}\d X
-{1\ov \mu_{i}}{\bf s}_{i}\d \psi^{*}, \
G^{-}_{i}={\bf s}^{*}_{i}\psi {\bf s}_{i}\d X^{*}
-{\bf s}^{*}_{i}\d \psi, \nmb
J_{i}=
({\bf s}_{i}\psi^{*}{\bf s}^{*}_{i}\psi+
{1\ov \mu_{i}}{\bf s}_{i}\d X^{*}-{\bf s}^{*}_{i}\d X), \nmb
T_{i}(z)=
({1\ov 2}({\bf s}_{i}\d \psi^{*}{\bf s}^{*}_{i}\psi-
{\bf s}_{i}\psi^{*}{\bf s}^{*}_{i}\d \psi)+
{\bf s}_{i}\d X^{*}{\bf s}^{*}_{i}\d X-  \nmb
{1\ov 2}({\bf s}^{*}_{i}\d^{2} X+{1\ov \mu_{i}}{\bf s}_{i}\d^{2} X^{*}))
\label{2.min}
\enr
as well as the vertex operators
\ber
V_{(p_{i},p^{*}_{i})}=
\exp(p^{*}_{i}{\bf s}^{*}_{i}X+p_{i}{\bf s}_{i}X^{*})),
\label{2.V}
\enr
which are the conformal fields:
\ber
G^{+}_{i}(z_{1})V_{(p_{i},p^{*}_{i})}(z_{2})=
z_{12}^{-1}p_{i}{\bf s}_{i}\psi^{*}V_{(p_{i},p^{*}_{i})}(z_{2})+r.,\
G^{-}_{i}(z_{1})V_{(p_{i},p^{*}_{i})}(z_{2})=
z_{12}^{-1}p^{*}_{i}{\bf s}^{*}_{i}\psi V_{(p_{i},p^{*}_{i})}(z_{2})+r.,\nmb
J_{i}(z_{1})V_{(p_{i},p^{*}_{i})}(z_{2})=
z_{12}^{-1}{1\ov \mu_{i}}j_{i}
V_{(p_{i},p^{*}_{i})}(z_{2})+r, \nmb
T_{i}(z_{1})V_{(p_{i},p^{*}_{i})}(z_{2})=
z_{12}^{-2}{1\ov 4\mu_{i}}(h_{i}(h_{i}+2)-j_{i}^{2})
V_{(p_{i},p^{*}_{i})}(z_{2})+ \nmb
z_{12}^{-1}\d V_{(p_{i},p^{*}_{i})}(z_{2})+r.,
\label{2.conf}
\enr
where
\ber
h_{i}=p^{*}_{i}+\mu_{i} p_{i}, \
j_{i}=p^{*}_{i}-\mu_{i} p_{i}.
\label{2.lm}
\enr

 These vertex operators are naturally associated to the
lattice $\Pi=P\oplus P^{*}\in E\oplus E^{*}$,
where $P\in E, P^{*}\in E^{*}$ such that
$P$ is generated by ${1\ov \mu_{i}}{\bf s}_{i}$ and
$P^{*}$ is generated by the basis
${\bf s}^{*}_{i}$, $i=1,...,r$.
For an arbitrary vector $(\bp,\bp^{*})\in\Pi$,
we introduce in NS sector Fock vacuum state $|\bp,\bp^{*}>$ by the
formulas similar to (\ref{1.vac}) and denote by $F_{\bp,\bp^{*}}$ the Fock module
generated from $|\bp,\bp^{*}>$ by the creation operators of the fields
$X_{i}(z), X^{*}_{i}(z)$, $\psi_{i}(z), \psi^{*}_{i}(z)$.

 Let $F_{\Pi}$ be the direct sum of Fock modules associated to the lattice $\Pi$.
As an obvious generalization of the results
from Sec.1. we form for each vector $\bh=\sum_{i}h_{i}{\bf s}^{*}_{i}\in P^{*}$,
where $h_{i}=0,1,...,\mu_{i}-2$ butterfly resolution
$C^{\star}_{\bh}$ as the product $\otimes_{i=1}^{r}C^{\star}_{h_{i}}$ of butterfly resolutions
of minimal models.
The corresponding ghost number operator $g$ is given by the sum of
ghost number operators of each of the resolutions. The
differential $\d$ acting on ghost number
$N$ subspace of the resolution
is given by the sum of differentials of each of the complexes $C^{\star}_{h_{i}}$.
It is obvious that the complex $C^{\star}_{\bh}$ is exact
except at the $F_{0,\bh}$ module, where the cohomology is given by the product
$M_{\bh,0}=\otimes_{i=1}^{r}M_{h_{i},0}$ of the chiral modules
of each minimal model. Hence one can represent the character
\ber
\chi_{\bh,0}(q,u)\equiv Tr_{M_{\bh,0}}(q^{L[0]-{c\ov24}})u^{J[0]})
\label{2.char}
\enr
of $M_{\bh,0}$
as the product of characters $\chi_{\bh,0}(q,u)=\prod_{i}\chi_{h_{i},0}(q,u)$.

 According to the discussion at the end of Sec.1. we obtain the
resolution and character for the product
of arbitrary irreducible modules of minimal models acting on $C^{\star}_{\bh}$ by the
spectral flow operators $U^{-\bt}=\prod_{i}U_{i}^{-t_{i}}$ of the minimal
models. Hence one can label the resolutions, modules and characters by the pairs of vectors $(\bh,\bt)$, from the
set $\Dl'=\{(\bh,\bt)|h_{i}=0,...,[{\mu_{i}\ov2}]-1,\
t_{i}=0,...,\mu_{i}-1,\ i=1,...,r\}$. On the equal footing one can use the set
$\Dl=\{(\bh,\bt)|h_{i}=0,...,\mu_{i}-2,\
t_{i}=0,...,h_{i},\ i=1,...,r\}$. 

 It is also clear that R sector resolutions, modules and characters are generated
from NS sector by the spectral flow $U^{-\bv/2}=\prod_{i=1}^{r}U_{i}^{-1/2}$,
where $\bv=(1,...,1)$ is $r$-dimensional vector.

 The same free-field realization can be used in the
right-moving sector. Thus the sets of screening
vectors  $\bar{\Re}$ and $\bar{\Re}^{*}$ have to be fixed in the
right-moving sector. It can be done in many ways, the only
restriction is that the corresponding cohomology group has to be
isomorphic to the space of states of the product of minimal models in the
right-moving sector. Therefore $\bar{\Re}$ and $\bar{\Re}^{*}$
are determined modulo $O(r,r)$ transformations which left unchanged
the matrix of scalar products $<{\bf s}_{i},{\bf s}^{*}_{j}>$.
In what follows we
put
\ber
\bar{\Re}=\Re, \ \bar{\Re}^{*}=\Re^{*}.
\label{2.bbases}
\enr
Hence, one can use the
same complex to describe the irreducible modules in the
right-moving sector.

\leftline {\bf 2.2.Free-field realization and Calabi-Yau extension.}

 It is well known that product of minimal models can not be
applied straightforward to describe the string theory on
$2D$-dimensional CY manifold. First, one has to demand
that $\sum_{i}c_{i}=3D$. Second, the so called simple current
orbifold ~\cite{SCORB} of the product of minimal models has to be constructed.
The orbifold, which is known as CY
extension ~\cite{EOTY}, ~\cite{FSW}, gives the space of states of
$N=2$ superconformal sigma model on CY model. We denote this model
by $CY_{\bmu}$. The currents
of $N=2$ Virasoro superalgebra of this model are given by the sum of
currents of each minimal model
\ber
G^{\pm}(z)=\sum_{i}G^{\pm}_{i}, \nmb
J(z)=\sum_{i}J_{i}, \
T(z)=\sum_{i}T_{i}.
\label{2.Vird}
\enr

 The left-moving (as well as the right-moving) sector of the $CY_{\bmu}$ is given by
projection of the space of states on the
subspace of integer $J[0]$-charges and organizing the projected
space into the orbits $[\bh,\bt]$ under the spectral flow operator
$U^{\bv}=\prod_{i=1}^{r}U_{i}$ ~\cite{FSW}.

 The partition function in NS sector of $CY_{\bmu}$ sigma model
is diagonal modular invariant of the spectral flow orbits characters restricted to the
subset of integer $J[0]$ charges. From $N=2$ Virasoro superalgebra representations there is no
difference what of the sets $\Dl$ or $\Dl'$ we use to parameterize the orbit characters (though their
free-field realizations are different). In what follows
we combine these to sets into the
extended set $\tld{\Dl}=\{(\bh,\bt)|h_{i}=0,...,\mu_{i}-2,\
t_{i}=0,...,\mu_{i}-1,\ i=1,...,r\}$ which is $2^{r}$ times bigger than $\Dl$ or $\Dl'$
and take into account this extension by corresponding multiplier ("field identification")
~\cite{Gep}.

 The orbit characters (with the restriction on integer charges
subspace) can be written in explicit form so that the structure of simple
current extension becomes clear ~\cite{EOTY},~\cite{FSW}:
\ber
ch_{\bh,\bt}(q,u)={1\ov\kp^{2}}\sum_{n,m=0}^{\kp-1}
Tr_{M_{\bh,\bt}}(U^{n\bv}q^{(L[0]-{c\ov24})}u^{J[0]}\exp{(\im2\pi
mJ[0])}U^{-n\bv})=\nmb
{1\ov\kp^{2}}\sum_{n,m=0}^{\kp-1}\exp{(\im2\pi({cn^{2}\ov6}\tau+{cn\ov3}\ups))}
\chi_{\bh,\bt}(\tau,\ups+n\tau+m)=\nmb
{1\ov\kp^{2}}\sum_{n,m=0}^{\kp-1}\chi_{\bh,\bt+n\bv}(\tau,\ups+m),
\label{2.orbchi}
\enr
where $q=\exp{(\im2\pi\tau)}$, $u=\exp{(\im2\pi\ups)}$ and
$\kp=lcm\{\mu_{i}\}$.
The partition function of $CY_{\bmu}$ sigma model is given by
\ber
Z_{CY}(q,\bar{q})={1\ov 2^{r}}\sum_{[\bh,\bt]\in \Dl_{CY}}\kp|ch_{[\bh,\bt]}(q)|^{2},
\label{2.ZCY}
\enr
where $\Dl_{CY}$ denotes the subset of $\tld{\Dl}$ restricted to the space of integer $J[0]$ charges.
$[\bh,\bt]$ denotes the spectral flow orbit of the point $(\bh,\bt)$.
Factor ${1\ov 2^{r}}$ corresponds to the extended set $\tld{\Dl}$ of irreducible modules and
$\kp$ is the length of the orbit $[\bh,\bt]$. In general case the
orbits with different lengths could appear but we will not consider these
cases to escape the problem of fixed point resolution ~\cite{SCORB}, ~\cite{FSW},
~\cite{BRS}.
 Modular invariance of the partition function is due to the following
behavior under modular transformation ~\cite{Gep},~\cite{FSW}
\ber
ch_{[\bt,\bh]}(-{1\ov\tau})=\kp\sum_{[\bh',\bt]\in
\Dl_{CY}}S_{[\bh,\bt],[\bh',\bt']}ch_{[\bt',\bh']}(\tau),
\label{2.Strans}
\enr
where the matrix $S_{[\bh,\bt],[\bh',\bt']}$ is given by the
product of modular $S$ matrices of minimal models factors.

\leftline {\bf 2.3.Free-field realization of Gepner
models.}

 The Gepner models ~\cite{Gep} of CY superstring
compactification are given by (generalized) GSO projection ~\cite{Gep}, ~\cite{EOTY}
which is carrying out on the product of the space of states
of $CY_{\bmu}$ $\sigma$-model and space of states of external fermions and
bosons describing space-time degrees of freedom of the string.
In the framework of simple current extension formalism the Gepner's construction
has been farther developed in ~\cite{FSW},
~\cite{FSS}, ~\cite{SCORB}.

 Let us introduce so called supersymmetrized (Green-Schwartz)
characters ~\cite{Gep},~\cite{EOTY}
\ber
Ch_{[\bh,\bt]}(q,u)={1\ov 4\kp^{2}}\sum_{n,m=0}^{2\kp-1}
Tr_{(M_{\bh,\bt}\otimes\Phi)}(U_{tot}^{m\ov2}\exp{(\im\pi nJ_{tot}[0])}
q^{(L_{tot}-{c_{tot}\ov24})}u^{J_{tot}[0]}U_{tot}^{-m\ov2}),
\label{3.gschar}
\enr
where the trace is calculated in the product of
$M_{\bh,\bt}$ and Fock module $\Phi$
generated by the external (space-time) fermions and bosons in NS sector,
$J_{tot}[0]$ and $L_{tot}[0]$ are zero modes of the total $U(1)$ current
and stress-energy tensor which includes the contributions from space-time
degrees of freedom, $c_{tot}=c+{3\ov2}(8-2D)=12$ is a total central charge and
$U_{tot}$ is a total spectral flow operator acting
in the product $M_{\bh,\bt}\otimes\Phi$.

 The modular invariant Gepner model partition function is given by
~\cite{Gep},~\cite{EOTY},~\cite{FSW}
\ber
Z_{Gep}(q,\bar{q})={1\ov2^{r}}(Im\tau)^{-(4-D/2)}
\sum_{[\bh,\bt]\in \Dl_{CY}}\kp|Ch_{[\bh,\bt]}(q)|^{2}.
\label{3.ZG}
\enr

\vskip 10pt
\centerline{\bf 3. Free-field representation for Ishibashi states in}
\centerline{\bf Gepner models.}
\vskip 10pt
\leftline {\bf 3.1. Linear Ishibashi states in Fock modules.}

 The boundary states we are going to construct can be considered
as a bilinear forms on the space of states of the model. Thus, it
will be implied in what follows that the right-moving sector of
the model is realized by the free-fields
$\bar{X}_{i}(\bar{z}), \bar{X}^{*}_{i}(\bar{z}), \bar{\psi}_{i}(\bar{z}),
\bar{\psi}^{*}_{i}(\bar{z})$, $i=1,...,r$ and the right-moving
$N=2$ super-Virasoro algebra is given by the formulas similar to
(\ref{1.min})

 There are two types of boundary states preserving $N=2$
super-Virasoro algebra ~\cite{OOY}, usually called $B$-type
\ber
(L[n]-\bar{L}[-n])|B>>=(J[n]+\bar{J}[-n])|B>>=0, \nmb
(G^{+}[r]+\imath \et \bar{G}^{+}[-r])|B>>=
(G^{-}[r]+\imath \et \bar{G}^{-}[-r])|B>>=0
\label{3.BD}
\enr
and $A$-type states
\ber
(L[n]-\bar{L}[-n])|A>>=(J[n]-\bar{J}[-n])|A>>=0, \nmb
(G^{+}[r]+\im \et \bar{G}^{-}[-r])|A>>=
(G^{-}[r]+\im \et \bar{G}^{+}[-r])|A>>=0
\label{3.AD}
\enr
where $\et=\pm 1$.

 In the tensor product
of the left-moving Fock module $F_{\bp,\bp^{*}}$ and right-moving Fock
module $\bar{F}_{\bar{\bp},\bar{\bp}^{*}}$ we construct the most
simple states fulfilling the solutions (\ref{3.BD}) and
(\ref{3.AD}). We shall call these states as Fock space or linear Ishibashi~\cite{Ish}
states and denote by $|\bp,\bp^{*},\bar{\bp},\bar{\bp}^{*},\et,B(A)>>$.

 We consider first $B$-type linear Ishibashi states. In NS sector
they can be easily obtained from the following ansatz for fermions
\ber
(\psi^{*}_{i}[r]-
\im \et \bar{\psi}^{*}_{i}[-r])|\bp,\bp^{*},\bar{\bp},\bar{\bp}^{*},\et,B>>=0, \nmb
(\psi_{i}[r]-
\im \et \bar{\psi}_{i}[-r])|\bp,\bp^{*},\bar{\bp},\bar{\bp}^{*},\et,B>>=0.
\label{3.Bantz}
\enr
Substituting these relations
into (\ref{3.BD}) and using (\ref{2.min}), (\ref{2.Vird}) we find
\ber
(X_{k}[n]+\bar{X}_{k}[-n]+d_{k}\dlt_{n,0})
|\bp,\bp^{*},\bar{\bp},\bar{\bp}^{*},\et,B>>=0, \nmb
(X^{*}_{k}[n]+\bar{X}^{*}_{k}[-n]+d^{*}_{k}\dlt_{n,0})
|\bp,\bp^{*},\bar{\bp},\bar{\bp}^{*},\et,B>>=0,
\label{3.BX}
\enr
where $d_{k}={1\ov\mu_{k}}$, $d^{*}_{k}=1$ and we combine these coefficients into
the $r$-dimensional vectors $\bd=(d_{1},...,d_{r})$, $\bd^{*}=(d^{*}_{1},...,d^{*}_{r})$.
It follows from these relations that we have $B$-type boundary
conditions for each minimal model.

The linear $B$-type Ishibashi state in NS sector is given by the
standard expression ~\cite{CLNY},~\cite{PC}
\ber
|\bp,\bp^{*},\et,B>>=
\prod_{n=1}\exp(-{1\ov
n}\sum_{i}(X^{*}_{i}[-n]\bar{X}_{i}[-n]+
X_{i}[-n]\bar{X}^{*}_{i}[-n])) \nmb
\prod_{r=1/2}\exp(\im\et\sum_{i}(\psi^{*}_{i}[-r]\bar{\psi}_{i}[-r]+
\psi_{i}[-r]\bar{\psi}^{*}_{i}[-r]))
|\bp,\bp^{*},-\bp-\bd,-\bp^{*}-\bd^{*}>.
\label{3.BI}
\enr

 The closed string transition amplitude between such states
in NS sector
is given by
\ber
<<\bp_{2},\bp_{2}^{*},\et,B|q^{L[0]-c/24}u^{J[0]}
|\bp_{1},\bp_{1}^{*},\et,B>>=\nmb
\dlt(\bp_{1}-\bp_{2})
\dlt(\bp_{1}^{*}-\bp_{2}^{*})
q^{{1\ov2}<\bp_{1}+\bp^{*}_{1},\bp_{1}+\bp^{*}_{1}+\bd+\bd^{*}>-{c\ov24}}
u^{<\bp_{1}+\bp^{*}_{1},\bd-\bd^{*}>}\nmb
\prod_{m=1}(1+uq^{m-{1\ov2}})^{r}
(1+u^{-1}q^{m-{1\ov2}})^{r}
(1-q^{m})^{-2r}.
\label{3.Bampl}
\enr
Note that the state $<<\bp,\bp^{*},\et,B|$ is defined in such
a way to satisfy conjugate boundary conditions and to take into account
the charge asymmetry ~\cite{FeFu}-~\cite{Fel}
of the free-field realization of each minimal model.

The linear $A$-type Ishibashi states can be found analogously. The
linear ansatz for fermions has the form
\ber
(\psi^{*}_{i}[r]-
\im \et \mu_{i}\bar{\psi}_{i}[-r])
|\bp,\bp^{*},\bar{\bp},\bar{\bp}^{*},\et,A>>=0, \nmb
(\psi_{i}[r]-
{\im \et \ov\mu_{i}}\bar{\psi}^{*}_{i}[-r])
|\bp,\bp^{*},\bar{\bp},\bar{\bp}^{*},\et,A>>=0.
\label{3.Aantz}
\enr
Substituting these relations
into (\ref{3.AD}) and using (\ref{1.min}) we find
\ber
(\mu_{k}X_{k}[n]+\bar{X}^{*}_{k}[-n]+d_{k}^{*}\dlt_{n,0})
|\bp,\bp^{*},\bar{\bp},\bar{\bp}^{*},\et,A>>=0, \nmb
({1\ov\mu_{k}}X^{*}_{k}[n]+\bar{X}_{k}[-n]+d_{k}\dlt_{n,0})
|\bp,\bp^{*},\bar{\bp},\bar{\bp}^{*},\et,A>>=0.
\label{3.AX}
\enr
Therefore we have $A$-type boundary
conditions for each minimal model.

 The linear $A$-type Ishibashi state in NS sector is given similar
to $B$-type
\ber
|\bp,\bp^{*},\et,A>>=
\prod_{n=1}\exp(-{1\ov
n}\sum_{i}(\mu_{i}X_{i}[-n]\bar{X}_{i}[-n]+
{1\ov\mu_{i}}X^{*}_{i}[-n]\bar{X}^{*}_{i}[-n])) \nmb
\prod_{r=1/2}\exp(\im\et\sum_{i}(\mu_{i}\psi_{i}[-r]\bar{\psi}_{i}[-r]+
{1\ov\mu_{i}}\psi^{*}_{i}[-r]\bar{\psi}^{*}_{i}[-r]))
|\bp,\bp^{*},-\Om^{-1}\bp^{*}-\bd,-\Om\bp-\bd^{*}>,
\label{3.AI}
\enr
where we have introduced the matrix $\Om_{ij}=\mu_{i}\dlt_{ij}$.

 The closed string transition amplitude between $A$-type Ishibashi states
in NS sector is given by
\ber
<<\bp_{2},\bp_{2}^{*},\et,A|q^{L[0]-c/24}u^{J[0]}|\bp_{1},\bp_{1}^{*},\et,A>>=\nmb
\dlt(\bp_{1}-\bp_{2})
\dlt(\bp_{1}^{*}-\bp_{2}^{*})
q^{{1\ov2}<\bp_{1}+\bp^{*}_{1},\bp_{1}+\bp^{*}_{1}+\bd+\bd^{*}>-{c\ov24}}
u^{<\bp_{1}+\bp^{*}_{1},\bd-\bd^{*}>}\nmb
\prod_{m=1}(1+uq^{m-{1\ov2}})^{r}
(1+u^{-1}q^{m-{1\ov2}})^{r}
(1-q^{m})^{-2r}.
\label{3.Aampl}
\enr

\vskip 10pt
\leftline {\bf 3.2. $B$-type Ishibashi states in the product of minimal models.}
  
 Free-field realizations of the irreducible modules described in Sect. 1,2
and the constructions (\ref{3.BI}), (\ref{3.AI})
allows to suggest that Ishibashi states can also be represented by the free-fields.
Let us consider the following superposition of $B$-type free-fields 
Ishibashi states (\ref{3.BI})
\ber
|I_{\bh},\et,B>>=\sum_{(\bp,\bp^{*})\in\Gm_{\bh}}c_{\bp,\bp^{*}}|\bp,\bp^{*},\et,B>>,
\label{3.Isuper}
\enr
where the coefficients $c_{\bp,\bp^{*}}$ are arbitrary and the summation is
performed over the momentums of the butterfly resolution
$C^{\star}_{\bh}$. It is clear that this state satisfies the relations
(\ref{3.BD}). We define the closed string transition amplitude between the pair
of such states by the following expression $<<I_{\bh_{2}},\et,B|(-1)^{g}q^{L[0]-{c\ov
24}}u^{J[0]}|I_{\bh_{1}},\et,B>>$, where $g$ is ghost number operator of the complex $C^{\star}_{\bh}$.
It is easy to see that
\ber
<<I_{\bh_{2}},\et,B|(-1)^{g}q^{L[0]-{c\ov
24}}u^{J[0]}|I_{\bh_{1}},\et,B>>=\nmb
\dlt(\bh_{1}-\bh_{2})
\sum_{(\bp,\bp^{*})\in\Gm_{\bh_{1}}}(-1)^{g}|c_{\bp,\bp^{*}}|^{2}\nmb
q^{{1\ov2}<\bp^{*}+\bp,\bp^{*}+\bp+\bd+\bd^{*}>-{c\ov24}}
u^{<\bp^{*}+\bp,\bd-\bd^{*}>}\nmb
\prod_{m=1}(1+uq^{m-{1\ov2}})^{r}
(1+u^{-1}q^{m-{1\ov2}})^{r}
(1-q^{m})^{-2r}.
\label{3.IBamp}
\enr
 The coefficients $c_{\bp,\bp^{*}}$ can be fixed partly
from the condition that this amplitude is proportional to the character of the
module
$M_{\bh}$:
\ber
<<I_{\bh_{2}},\et,B|(-1)^{g}q^{L[0]-{c\ov
24}}u^{J[0]}|I_{\bh_{1}},\et,B>>=\dlt(\bh_{2}-\bh_{1})
|c_{0,\bh_{1}}|^{2}\chi_{\bh_{1}}(q,u).
\label{3.norm}
\enr
Comparing with (\ref{1.char}) we obtain
\ber
|c_{\bp,\bp^{*}}|^{2}=|c_{0,\bh}|^{2}, \ (\bp,\bp^{*})\in\Gm_{\bh}.
\label{3.modc}
\enr
Thus, the state (\ref{3.Isuper}) is a good candidate for free-field
realization (in NS sector) of $B$-type Ishibashi state of the
module $M_{\bh}$. It would be a genuine Ishibashi state if it did
not radiate nonphysical closed string states which are present in the
free-field representation of the model. In other words, the overlap of
this state with an
arbitrary closed string state which does not belong to
the Hilbert space of the $MM_{\bmu}$ model should vanish. As we will see
this condition can be formulated as a $BRST$ invariance condition
of the state (\ref{3.Isuper}).

 Before GSO projection, closed string states of $MM_{\bmu}$ model
which can interact with the Ishibashi state
(\ref{3.Isuper}) come from the product of left-moving and right-moving
Fock modules
$F_{\bp,\bp^{*}}\otimes
\bar{F}_{-\bp-\bd,-\bp^{*}-\bd^{*}}$, where
$(\bp,\bp^{*})\in\Gm_{\bh}$. The left-moving modules of the superposition
(\ref{3.Isuper}) constitute the
butterfly resolution $C^{\star}_{\bh}$ whose cohomology is given by
the module $M_{\bh}$. What about the Fock modules from
the right-moving sector? To have nontrivial interaction with the states
from $MM_{\bmu}$ the right-moving Fock modules
have to from the product of resolutions of minimal models
(\ref{1.but}) also. But this contradicts to the relations between
left-moving and right-moving momentums from (\ref{3.BX}).  The
contradiction may be resolved if we allow the right-moving Fock
modules to form the product of dual resolutions
to (\ref{1.but}). The dual resolution $\tld{C}^{\star}_{h}$ to minimal model
resolution (\ref{1.but})
is given by the following diagram
\ber
\begin{array}{ccccccccccc}
&&\vdots &\vdots &&&&&&\\
&&\downarrow &\downarrow &&&&&&\\
\ldots &\rightarrow &\bar{F}_{-1-{1\ov\mu},-1-h-\mu} &\rightarrow
\bar{F}_{-{1\ov\mu},-1-h-\mu}&&&&&&\\
&&\downarrow &\downarrow &&&&&&\\
\ldots &\rightarrow &\bar{F}_{-1-{1\ov\mu},-1-h} &\rightarrow
\bar{F}_{-{1\ov\mu},-1-h}&&&&&&\\
&&&&\searrow&&&&&\\
&&&&&\bar{F}_{1-{1\ov\mu},-1-h+\mu}&\rightarrow &\bar{F}_{2-{1\ov\mu},-1-h+\mu}
&\rightarrow&\ldots\\
&&&&&\downarrow &&\downarrow&\\
&&&&&\bar{F}_{1-{1\ov\mu},-1-h+2\mu}&\rightarrow &\bar{F}_{2-{1\ov\mu},-1-h+2\mu}
&\rightarrow&\ldots\\
&&&&&\downarrow &&\downarrow &&\\
&&&&&\vdots &&\vdots &&
\end{array} \nmb
\label{3.dualbut}
\enr

The arrows on this diagram are given by the same operators as on
the diagram (\ref{1.but}).

 Similar to the ~\cite{FeS} one can show that complex (\ref{3.dualbut}) is exact
except at the $\bar{F}_{-{1\ov\mu},-1-h}$ module, where the cohomology is isomorphic
to the anti-chiral module $M_{h,t=2h}$.

 Thus the right-moving Fock modules have to form dual resolution
$\tld{C}^{\star}_{\bh}=\otimes_{i=1}^{r}\tld{C}^{\star}_{h_{i}}$.
Now the coefficients $c_{\bp,\bp^{*}}$ can be defined from the $BRST$
invariance condition which is straightforward generalization of the condition
found in ~\cite{SP} for individual $N=2$ minimal model. To formulate this
condition one has to describe by the free-fields the total space
of states of $MM_{\bmu}$ model.

 To do that we form first the product of complexes $C^{\star}_{\bh}\otimes
\tld{C}^{\star}_{\bh}$ to build the complex
\ber
\ldots\rightarrow {\bf C}_{\bh}^{-2}\rightarrow {\bf C}_{\bh}^{-1}
\rightarrow {\bf C}_{\bh}^{0}\rightarrow {\bf C}_{\bh}^{+1}\rightarrow\ldots,
\label{3.complex}
\enr
which is graded by the sum of the ghost numbers
$g+\bar{g}$ and for an arbitrary ghost number $I$
the space ${\bf C}_{\bh}^{I}$ is given by the sum of products of the
Fock modules from the resolution $C^{\star}_{\bh}$ and $\tld{C}^{\star}_{\bh}$
such that $g+\bar{g}=I$. The differential $\dlt$ of the complex
(\ref{3.complex}) is defined by the differentials $\d$ and
$\bar{\d}$ of the complexes $C^{\star}_{\bh}$ and $\tld{C}^{\star}_{\bh}$
\ber
\dlt|v_{g}\otimes\bar{v}_{\bar{g}}>=|\d v_{g}\otimes\bar{v}_{\bar{g}}>+
(-1)^{g}|v_{g}\otimes\bar{\d}\bar{v}_{\bar{g}}>,
\label{3.Diff}
\enr
where $|v_{g}>$ is an arbitrary vector from the complex $C_{h}$
with ghost number $g$, while $|\bar{v}_{\bar{g}}>$ is an
arbitrary vector from the complex $\tld{C}^{\star}_{h}$ with the ghost
number $\bar{g}$ and $g+\bar{g}=I$.
The cohomology
of the complex (\ref{3.complex}) is nonzero only at grading 0
and is given by the product of irreducible modules
$M_{\bh}\otimes\bar{M}_{\bh,\bt=2\bh}$, where $\bar{M}_{\bh,\bt=2\bh}$
is the product of anti-chiral modules of minimal models.

 The Ishibashi state we are looking for
can be considered as a linear functional on the Hilbert space of the product of
models, then it has to be an element
of the homology group. Therefore, the $BRST$ invariance
condition for the state can be formulated as follows.

 Let us define the action of the differential $\dlt$ on the state
$|I_{\bh},\et,B>>$ by the formula
\ber
<<\dlt^{*}I_{\bh},\et,B|v_{g}\otimes \bar{v}_{\bar{g}}>\equiv
<<I_{\bh},\et,B|\dlt_{g+\bar{g}}|v_{g}\otimes \bar{v}_{\bar{g}}>,
\label{3.D*}
\enr
where $v_{g}\otimes \bar{v}_{\bar{g}}$ is an arbitrary element
from ${\bf C}_{\bh}^{g+\bar{g}}$. Then, $BRST$ invariance condition
means that
\ber
\dlt^{*}|I_{\bh},\et,B>>=0.
\label{3.Dcycl}
\enr

 As a straightforward generalization of Theorem 2 from ~\cite{SP}
we find that superposition (\ref{3.Isuper}) satisfies
$BRST$ invariance condition (\ref{3.Dcycl}) if the coefficients
$c_{\bp,\bp^{*}}$ take values $\pm 1$ according to  the expression
\ber
c_{\bp,\bp^{*}}=\sqrt{2}\cos((2g_{\bp,\bp^{*}}+1){\pi\ov4})c_{0,\bh},
\label{3.c}
\enr
where $g_{\bp,\bp^{*}}$ is the ghost number.
 Note that $BRST$ condition doesn't fix the phase of the overall
coefficient $c_{0,\bh}$.

 For an arbitrary module $M_{\bh,\bt}$, $(\bh,\bt)\in \tld{\Dl}$,
the Ishibashi state is generated
by the action of spectral flow operators on the Ishibashi state
$|I_{\bh},\et,B>>$:
\ber
|I_{\bh,\bt},\et,B>>=
\prod_{i}U_{i}^{t_{i}}\bar{U}_{i}^{-t_{i}}
|I_{\bh},\et,B>>.
\label{3.IBhj}
\enr
It is easy to see from (\ref{1.3}) that this state satisfy the
boundary conditions (\ref{3.Bantz}), (\ref{3.BX}), (\ref{3.Bminm}). Hence
(\ref{3.BD}) is fulfilled. It is also $BRST$ closed because
the spectral flow commutes with screening charges and transition amplitude
between a pair of
such states is proportional to the character $\chi_{\bh,\bt}$.

%%%%%%%%%%%%%%%%%%%%%%%%%%%%%%%%%%%%%%%%%%%%%%%%%%%%%%%%%%%%%%%%%%%%%%%%%%%%
%%%%%%%%%%%%%%%%%%%%%%%%%%%%%%%%%%%%%%%%%%%%%%%%%%%%%%%%%%%%%%%%%%%%%%%%%%%%%
\vskip 10pt
\leftline {\bf 3.3. $A$-type Ishibashi states in the product of minimal models.}

 Let us consider free-field representation for $A$-type
Ishibashi states. It is obvious that $A$-type Ishibashi states are given
by superpositions like (\ref{3.Isuper}), where the coefficients
are determined by the relation (\ref{3.modc}).
The $BRST$ condition for $A$-type states is slightly different
from $B$-type case. The reason is that the application of one of the
left-moving $BRST$ charges, say $Q^{+}_{i}$ to $A$-type state
gives according to (\ref{3.Aantz}) and (\ref{3.AX}) the
right-moving $BRST$ charge $\bar{Q}^{-}$ multiplied by
$\mu_{i}$ as opposed to the $B$-type case.
In fact we are free to rescale arbitrary the right-moving $BRST$ charges
because it does not change the cohomology of the complex in the
right-moving sector and the cohomology of the total complex
(\ref{3.complex}).
Hence we define the right-moving $BRST$ charges in such a way to
cancel this effect
\ber
\bar{S}^{+}_{i}(\bar{z})=
{\im\et\ov\mu_{i}}{\bf s}_{i}\bar{\psi}^{*}\exp({\bf s}_{i}\bar{X}^{*})(\bar{z}), \nmb
\bar{S}^{-}_{i}(\bar{z})=
\im\et\mu_{i}{\bf s}^{*}_{i}\bar{\psi}\exp(\mu_{i}{\bf s}^{*}_{i}\bar{X})(\bar{z}), \nmb
\bar{Q}^{\pm}_{i}=\oint d\bar{z} \bar{S}^{\pm}_{i}(\bar{z}),
\label{3.abarscr}
\enr

 As a result $BRST$ invariant
$A$-type Ishibashi state $|I_{\bh,\bh},\et,A>>$ is given similar to
(\ref{3.Isuper}), (\ref{3.c}) and similar to $B$-type case the phase
of coefficient $c_{0,\bh}$ is arbitrary also. For an arbitrary module
$M_{\bh,\bt}$ we then obtain
\ber
|I_{\bh,\bt},\et,A>>=
\prod_{i}U_{i}^{-t_{i}}\bar{U}_{i}^{-t_{i}}
|I_{\bh},\et,A>>.
\label{3.IAhj}
\enr

 Similar to the $B$-type case we find that closed string transition amplitude between $A$-type Ishibashi
states is given by the character $\chi_{\bh,\bt}$.

 In conclusion of this section the following remark is in order. In case when some of the $\mu_{i}$
coincide one can consider generalization of (\ref{3.Bantz})-(\ref{3.BX}) or (\ref{3.Aantz})-(\ref{3.AX}),
where left-moving degrees of freedom are related to the right-moving by a matrix. Carrying out the similar
analysis one can show that this matrix has to be a permutation matrix of identical minimal models.
It gives free-field realization of permutation branes ~\cite{Re}.

%%%%%%%%%%%%%%%%%%%%%%%%%%%%%%%%%%%%%%%%%%%%%%%%%%%%%%%%%%%%%%%%%%%%%%%%%%%%
%%%%%%%%%%%%%%%%%%%%%%%%%%%%%%%%%%%%%%%%%%%%%%%%%%%%%%%%%%%%%%%%%%%%%%%%%%%%

\vskip 10pt
\centerline{\bf 4. Free-field realization of boundary states in}
\centerline{\bf Gepner model.}
\vskip 10pt
\leftline {\bf 4.1. A-type boundary states in Calabi-Yau extension.}

 It is easy to obtain the set of boundary states in the product of minimal models
applying to the free-field realized
Ishibashi states the formula found by Cardy ~\cite{C}.
As we have seen $BRST$ invariance fixes the Ishibashi states up to
the arbitrary constant $c_{\bh,\bt}$ and we put (following to normalization by
Cardy ~\cite{C}) these coefficients to be equal 1.

 As it has already been noticed the product of minimal models can not be
applied straightforward to describe in the bulk the string theory on
CY manifold. Instead, the so called simple current
orbifold whose
partition function is diagonal modular invariant partition function with
respect to orbit characters (\ref{2.orbchi})
describes.
The extension of this technique to the conformal field theory
with a boundary has been developed in ~\cite{ReS},~\cite{FSW},~\cite{BRS}, ~\cite{GJ}.
In our approach we follow mainly ~\cite{GJ}.

 A-type boundary states in CY extension are labeled by
spectral flow orbits $[\bLm,\blm]\in \tld{\Dl}$. We consider first spectral flow invariant boundary
states. Their
expansion with respect to the Ishibashi states (\ref{3.IAhj})
can
be written according to Cardy's formula
\ber
|[\bLm,\blm],\et,A>>={\al\ov\kp^{2}}
\sum_{(\bh,\bt)\in \tld{\Dl}}W^{\bh,\bt}_{\bLm,\blm}
\sum_{m,n=0}^{\kp-1}\exp{(\im 2\pi nJ[0])}U^{m\bv}\bar{U}^{m\bv}|I_{\bh,\bt},\et,A>>,
\label{4.DA}
\enr
where $W^{\bh,\bt}_{{\bf\Lm},{\blm}}$ are Cardy's coefficients
\ber
W^{\bh,\bt}_{{\bf\Lm},{\blm}}=R^{\bh}_{\bLm}
\exp{(\im\pi\sum_{i=1}{(\Lm_{i}-2\lm_{i})(h_{i}-2t_{i})\ov\mu_{i}})},\
R^{\bh}_{\bLm}=\prod_{i=1}^{r}R^{h_{i}}_{\Lm_{i}}, \nmb
R^{h_{i}}_{\Lm_{i}}={S_{\Lm_{i},h_{i}}\ov\sqrt{S_{0,h_{i}}}}, \
S_{\Lm_{i},h_{i}}={\sqrt{2}\ov\mu_{i}}\sin({\pi(\Lm_{i}+1)(h_{i}+1)\ov\mu_{i}}),
\label{4.Cardy}
\enr
and $\al$ is the normalization constant.
The summation over $n$ makes $J[0]$-projection, while summation over $m$ introduce
spectral flow twisted sectors. $J[0]$-projection in the closed string sector
corresponds to spectral flow action in the open string sector ~\cite{BRS}, while spectral flow
action in the closed string sector corresponds to $J[0]$-projection in the open
string sector ~\cite{GJ}.

 This state depends only on the spectral flow orbit class.
Moreover, the $J[0]$ integer charge restriction of the orbits $[\bLm,\blm]$ is necessary for the
self-consistency of the expression (\ref{4.DA}). Indeed, on can change the parameterization of
the states $(\bh,\bt)$ by spectral flow shift
$(\bh,\bt'+l\bv)$ and insert this into (\ref{4.DA}). Then we obtain that this state is proportional to itself
with the coefficient $\exp{(\im2\pi l\sum_{i}{\Lm_{i}-2\lm_{i}\ov\mu_{i}})}$. Hence
\ber
\exp{(\im2\pi l\sum_{i}{\Lm_{i}-2\lm_{i}\ov\mu_{i}})}=1
\label{4.consist}
\enr
has to be satisfied which is nothing else $J[0]$ integer charge restriction. It means obviously
that the boundary state (\ref{4.DA}) is spectral flow invariant.

 It is also consistent with "field identifications"
\ber
|[\bLm,\blm],\et,A>>=|[\Lm_{1},...,\mu_{i}-\Lm_{i}-2,...,\Lm_{r},\lm_{1},...,\lm_{i}-\Lm_{i}-1,...,\lm_{r}],\et,A>>,
\label{4.DArefl}
\enr
which corresponds to automorphisms (\ref{1.reflect}).

The transition amplitude between these boundary states is given by
\ber
Z^{A}_{[\bLm_{1},\blm_{1}][\bLm_{2},\blm_{2}]}(\tld{q})\equiv
<<[\bLm_{1},\blm_{1}],\et,A|(-1)^{g}q^{L[0]-{c\ov24}}|[\bLm_{2},\blm_{2}],\et,A>>=
\nmb
\al^{2}\sum_{\bh,\bt}\prod_{i}(N_{\Lm_{1,i},\Lm_{2,i}}^{h_{i}}
\dlt^{(2\mu_{i})}(\Lm_{2,i}-2\lm_{2,i}-\Lm_{1,i}+2\lm_{1,i}+h_{i}-2t_{i})+
\nmb
N_{\Lm_{1,i},\Lm_{2,i}}^{\mu_{i}-h_{i}-2} \dlt^{(2\mu_{i})}(\Lm_{2,i}-2\lm_{2,i}-\Lm_{1,i}+2\lm_{1,i}+h_{i}-2t_{i}-\mu_{i}))
ch_{\bh,\bt}(\tld{q})
\label{4.amp5}
\enr
This expression is obviously invariant with respect to automorphisms
(\ref{1.reflect}). It can be rewritten in more compact form ~\cite{ReS}:
\ber
Z^{A}_{[\bLm_{1},\blm_{1}][\bLm_{2},\blm_{2}]}(\tld{q})=
\nmb
2^{r}\al^{2}\sum_{\bh,\bt}N_{\bLm_{1},\bLm_{2}}^{\bh}
\dlt^{(2\mu_{i})}(\bLm_{2}-2\blm_{2}-\bLm_{1}+2\blm_{1}+\bh-2\bt)ch_{\bh,\bt}(\tau),
\label{4.amp7}
\enr
where we have used the automorphism property of characters (\ref{1.reflect}) and the
factor $2^{r}$ is
caused by the "field identification" (\ref{1.reflect}). The important feature of the
spectral flow
invariant boundary states is that the open string spectrum is $J[0]$ projected. Hence
in Gepner
models the open string spectrum between these states will be BPS.

 We fix the constant $\al$
\ber
\al=2^{-r}
\label{4.norm}
\enr

 The internal automorphism group of Gepner model allows to construct additional
boundary states. Namely one can use the operator
$\exp{(-\im2\pi\sum_{i}\phi_{i}J_{i}[0])}\in U(1)^{r}$ to generate new
boundary states. Let us consider the properties of the state
\ber
|[\bLm,\blm],\bphi,\et,A>>
\equiv\exp{(-\im2\pi\sum_{i}\phi_{i}J_{i}[0])}|[\bLm,\blm],\et,A>>.
\label{4.DAphi}
\enr
It satisfies the conditions similar to (\ref{3.AD}) except the
relations for fermionic fields
\ber
(G^{\pm}[r]+\im\et\sum_{i}\exp{(\pm\im2\pi\phi_{i})}\bar{G}^{\mp}_{i}[-r])
|[\bLm,\blm],\bphi,\et,A>>=0,
\nmb
(\psi_{i}^{*}[r]-\im\et\mu_{i}\exp{(\im2\pi\phi_{i})}\bar{\psi}_{i}[-r])
|[\bLm,\blm],\bphi,\et,A>>=0,
\nmb
(\psi_{i}[r]-\im{\et\ov\mu_{i}}\exp{(-\im2\pi\phi_{i})}\bar{\psi}^{*}_{i}[-r])
|[\bLm,\blm],\bphi,\et,A>>=0.
\label{4.Aphi}
\enr
This state does not invariant with respect to diagonal N=2 Virasoro algebra unless
\ber
\phi_{i}\in Z, \ i=1,...,r.
\label{4.Aphi1}
\enr
Hence the group $U(1)^{r}$ reduces to $Z^{r}$. It is worth to note that one can ignore
the case when all $\phi_{i}$ are half-integer because it can be canceled by the $\et\rightarrow -\et$
redefinition.
It is easy to see directly what kind of states we obtain by this way.
\ber
|[\bLm,\blm],\bphi,\et,A>>=
\nmb
{\al\ov 2^{r}\kp^{2}}
\sum_{(\bh,\bt)\in \tld{\Dl}}W^{\bh,\bt}_{[{\bf\Lm},{\bf\lm}]}
\sum_{m,n=0}^{\kp-1}\exp{(\im 2\pi nJ[0])}\exp{(-\im2\pi m\sum_{i}\phi_{i}{c_{i}\ov3})}U^{m\bv}\bar{U}^{m\bv}
\nmb
\exp{(-\im2\pi\sum_{i}\phi_{i}{h_{i}-2t_{i}\ov\mu_{i}})}|I_{\bh,\bt},\et,A>>=
\nmb
{\al\ov 2^{r}\kp^{2}}
\sum_{(\bh,\bt)\in \tld{\Dl}}R_{\bLm}^{\bh}
\exp{(\im\pi\sum_{i}{(\Lm_{i}-2\lm_{i}-2\phi_{i})(h_{i}-2t_{i})\ov\mu_{i}})}
\nmb
\sum_{m,n=0}^{\kp-1}\exp{(\im 2\pi nJ[0])}
\exp{(\im4\pi m\sum_{i}{\phi_{i}\ov\mu_{i}})}U^{m\bv}\bar{U}^{m\bv}
|I_{\bh,\bt},\et,A>>.
\label{4.DAphi1}
\enr
It is one of the states (\ref{4.DA}) if
\ber
2\sum_{i}{\phi_{i}\ov\mu_{i}} \in Z.
\label{4.perm}
\enr
In opposite case the new boundary states are generated by this action. One can see that
all Recknagel-Schomerus states ~\cite{ReS} are generated by this action providing the
following parameterization
of Recknagel-Shomerus boundary states ~\cite{BDLR}
\ber
|[\bLm,\blm],\et,A>>=\exp{(-\im2\pi\sum_{i}\lm_{i}J_{i}[0])}|[\bLm,0],\et,A>>.
\label{4.DAphi1}
\enr

 The transition amplitude between these states is given by
\ber
Z^{A}_{([\bLm_{1},\blm_{1}],([\bLm_{2},\blm_{2}],)}
(\tld{q})=
\nmb
{\al^{2}\ov\kp^{2}}\sum_{(\bh,\bt)\in \tld{\Dl}}
\prod_{i}(N_{\Lm_{1,i},\Lm_{2,i}}^{h_{i}}
\dlt^{(2\mu_{i})}(\Lm_{2,i}-2\lm_{2,i}-\Lm_{1,i}+2\lm_{1,i}+h_{i}-2t_{i})+
\nmb
N_{\Lm_{1,i},\Lm_{2,i}}^{\mu_{i}-h_{i}-2}
\dlt^{(2\mu_{i})}(\Lm_{2,i}-2\lm_{2,i}-\Lm_{1,i}+2\lm_{1,i}+h_{i}-2t_{i}-\mu_{i}))
\nmb
\sum_{m,n=0}Tr_{\bh,\bt}(U^{-n\bv}\exp{(\im2\pi m(J[0]+\sum_{i}{c_{i}\ov3}(\lm_{2,i}-\lm_{1,i}))}
\tld{q}^{(L[0]-{c\ov24})}U^{n\bv})
\label{4.cyamphi}
\enr
We see that $J[0]$ projection is shifted by
$\sum_{i}{c_{i}\ov3}(\lm_{2,i}-\lm_{1,i})$.  Hence the tachyon my appear between
such boundary states in Gepner model.

\vskip 10pt
\leftline {\bf 4.2. B-type boundary states in Calabi-Yau extension.}

 $B$-type boundary state
can couple only to "charge conjugate" parts
$M_{\bh,\bt}\otimes\bar{M}_{\bh,\bmu+\bh-\bt}$of the bulk Hilbert
space ($t_{i}$ are defined modulo $\mu_{i}$). If all components
of vector $\bmu$ are odd (short spectral flow orbits do not appear) it forces the
restriction on the set of Ishibashi states ~\cite{ReS}:
\ber
\bmu+\bh-\bt=l\bv+\bt,
\label{4.chargconj}
\enr
where we have identified the vector $\bh$ with
$(h_{1},...,h_{r})$ and $l=0,...,\kp-1$.

Let us denote by $\tld{\Dl}_{B}$ the subset of
$\tld{\Dl}$ satisfying (\ref{4.chargconj}). Then
for an arbitrary pair of vectors $[\bLm,\blm]\in \Dl_{CY}$ the
ansatz for $B$-type boundary state is given by
\ber
|[\bLm,\blm],\et,B>>={\al\ov\kp^{2}}
\sum_{(\bh,\bt)\in {\tld{\Dl}_{B}}}W^{\bh,\bt}_{[{\bf\Lm},{\bf\lm}]}
\sum_{m,n=0}^{\kp-1}\exp{(\im 2\pi nJ[0])}
U^{m\bv}\bar{U}^{-m\bv}|I_{\bh,\bt},\et,B>>.
\label{4.DB}
\enr
One can check that this state depends only on the spectral flow orbit. It is also obvious that
$[\bLm,\blm]$ has to be restricted to the set of $J[0]$ integer charges by the reasons
similar to the $A$-type case.

 The transition amplitude calculation between the pair of $B$-type states can be carried over
similar to ~\cite{ReS}
\ber
Z^{B}_{[\bLm_{1},\blm_{1}][\bLm_{2},\blm_{2}]}(\tld{q},\tld{u})=
\nmb
2^{r}\kp\al^{2}
\sum_{(\bh,\bt)\in \tld{\Dl}_{CY}}
\dlt^{(\kp)}(\kp\sum_{i}{\Lm_{2,i}-2\lm_{2,i}-\Lm_{1,i}+2\lm_{1,i}+h_{i}'-2t_{i}'\ov2\mu_{i}})
N_{\bLm_{1},\bLm_{2}}^{\bh}ch_{\bh,\bt}(\tau).
\label{4.Btramp9}
\enr
Thus one has to put
\ber
\al=2^{-r}\kp^{-1}
\label{4.Bnorm}
\enr
to take into account that extended set $\tld{\Dl}$ has been used.

 Similar to the $A$-type case the open string spectrum between spectral flow invariant states
is $J[0]$ projected. Hence they give BPS spectrum in Gepner model.

 Similar to the $A$-type case the discreet group of internal automorphisms is acting on
the $B$-type boundary states and all Recknagel-Shomerus states can be recovered by this action
\ber
|[\bLm,\blm],\et,B>>
\equiv\exp{(\im\sum_{i}2\pi\lm_{i}J_{i}[0])}|[\bLm,0],\et,B>>,
\label{4.DBphi}
\enr
where $\lm_{i}$ are integers restricted to the region $\lm_{i}=0,...,\mu_{i}$.

 Closed string transition amplitude is given by
\ber
Z^{B}_{([\bLm_{1},\blm_{1}])([\bLm_{2},\blm_{2}])}(\tld{q})=
\nmb
\al^{2}\kp
\sum_{(\bh,\bt)\in \tld{\Dl}}
N_{\bLm_{1},\bLm_{2}}^{\bh}
\dlt^{(\kp)}(\kp
\sum_{i}{(\bLm_{2}-2\blm_{2}-\bLm_{1}+2\blm_{1}+\bh-2\bt)_{i}
\ov2\mu_{i}})
\nmb
\sum_{m,n}Tr_{M_{\bh'',\bt''}}(U^{-n\bv}\exp{(\im2\pi m(J[0]+ 
\sum_{i}{c_{i}\ov3}(\lm_{2,i}-\lm_{1,i}))})
\tld{q}^{(L[0]-{c\ov24})}U^{n\bv}).
\label{4.Btramphi}
\enr
Similar to the $A$-type case we see that $J[0]$ projection is shifted by
of $\sum_{i}{c_{i}\ov3}(\lm_{2,i}-\lm_{1,i})$. Hence in Gepner model the tachyon is expected in
the open string spectrum.

\vskip 10pt
\leftline {\bf 4.4. A-type boundary states in Gepner models.}

 In this subsection we discuss free-field construction of boundary
states in Gepner models using the free-field realization of
Ishibashi states we developed in section 3 and taking into account
GSO projection.

In Gepner model
partition function the supersymmetrized characters $Ch_{\bh,\bt}$
appear. Hence the natural idea ~\cite{ReS}, ~\cite{GJ} is to combine Cardy's
prescription and
supersymmetrization procedure to built the boundary states. In the work ~\cite{GJ}
the relationship between spacetime supersymmetry and spectral flow
has been straightforwardly used in BPS boundary state construction.
The representation of this section follows mainly the way of ~\cite{GJ}.

 Let us introduce the notation $|\Im_{\bh,\bt},\et,A>>$ for the
product of internal Ishibashi state $|I_{\bh,\bt},\et,A>>$ and
external Ishibashi state of Fock module $\Phi$. In this notation we omit (temporarily)
the indexes labeling external Ishibashi states.
 Let us consider first the following
ansatz for spectral flow invariant $A$-type boundary states in Gepner model (they are
still Ishibashi states in space-time sector)
\ber
|[\bLm,\blm],\et,A>>=
{\al\ov4\kp^{2}}\sum_{(\bh,\bt)\in
\tld{\Dl}}\sum_{m,n=0}^{2\kp-1}(-1)^{n}\exp{(\im\pi{m^{2}\ov2})}\nmb
W_{(\bLm,\blm)}^{(\bh,\bt)}
\exp{(\im\pi nJ_{tot}[0])}
U^{m\ov2}_{tot}\bar{U}^{m\ov2}_{tot}|\Im_{\bh,\bt},\et,A>>,
\label{4.ADgep}
\enr
where $\al$ is given by (\ref{4.norm}) and $W_{(\bLm,\blm)}^{(\bh,\bt)}$ is given by
(\ref{4.Cardy}).
The summation over $n$
gives the projection on the odd $J_{tot}[0]$ charges.
The Ramound sector contribution is included in the summation over $m$ such that
it comes with the coefficient $\im$.

The transition amplitude is given by
\ber
Z^{A}_{[\bLm_{1},\blm_{1}][\bLm_{2},\blm_{2}]}(\tld{q})=
<<([\bLm_{1},\blm_{1}],\et)^{*},A|(-1)^{g}q^{(L_{tot}[0]-{c_{tot}\ov24})}
|[\bLm_{2},\blm_{2}],\et,A>>,
\label{4.Dgepz}
\enr
where $|([\bLm_{1},\blm_{1}],\et)^{*},A>>$ is $CPT$ conjugated
state ~\cite{ReS}.
Using (\ref{4.ADgep}) we obtain
\ber
Z^{A}_{[\bLm_{1},\blm_{1}],[\bLm_{2},\blm_{2}]}(\tld{q},\tld{u})=
\nmb
(-\im\tld{\tau})^{D-4}\sum_{\bh,\bt}
N_{\bLm_{1},\bLm_{2}}^{\bh}
\dlt^{(2\mu_{i})}(\bLm_{2}-2\blm_{2}-\bLm_{1}+2\blm_{1}+\bh-2\bt)
Ch_{\bh,\bt}(\tld{q},\tld{u}),
\label{4.Dgepamp}
\enr
where the factor $(-\im\tld{\tau})^{D-4}$ is caused by modular transformation of spacetime bosons.
Because of supersymmetrized characters appears on the right-hand side the open string spectrum is
BPS.

 The other boundary states can be generated similar to (\ref{4.DAphi}).
For these boundary states it is easy to perform the calculation
similar to (\ref{4.Dgepamp}) and see that tachyon decouples from the open string
spectrum unless ~\cite{ReS}
\ber
\sum_{i}{c_{i}\ov3}(\lm_{2,i}-\lm_{1,i})\in Z.
\label{4.bps}
\enr

\vskip 10pt
\leftline {\bf 4.5. B-type boundary states in Gepner models.}

 The ansatz for $B$-type boundary states could be taken in the
form similar to $A$-type case. Obviously the Ishibashi
states that contribute to the superposition are restricted by
the subsets $\Dl_{B}' \in \Dl$ which is defined by the relation similar to
(\ref{4.chargconj}):
\ber
\bmu+\bh-(\bt-{m\ov 2}\bv)=(\bt-{m\ov 2}\bv)+{n\ov 2}\bv,
\label{4.gepconj}
\enr
where $m,n$ are defined modulo $2\kp$.
The ansatz for $B$-type spectral flow invariant boundary states is
\ber
|[\bLm,\blm],\et,B>>=
{\al\ov4\kp^{2}}\sum_{(\bh,\bt)\in
\tld{\Dl}_{B}'}\sum_{m,n=0}^{2\kp-1}(-1)^{n}\exp{(\im\pi{m^{2}\ov2})}\nmb
W_{(\bLm,\blm)}^{(\bh,\bt)}
\exp{(\im\pi nJ_{tot}[0])}
U^{m\ov2}_{tot}\bar{U}^{m\ov2}_{tot}|\Im_{\bh'\bt'},\et,B>>.
\label{4.BDgep}
\enr

 The transition amplitude is given by
\ber
Z^{B}_{[\bLm_{1},\blm_{1}][\bLm_{2},\blm_{2}]}(\tld{q})=
\nmb
(-\im\tld{\tau})^{D-4}
\sum_{(\bh,\bt)\in \tld{\Dl}}
N_{\bLm_{1},\bLm_{2}}^{\bh}
\dlt^{(\kp)}(\kp\sum_{i}{(\bLm_{2}-\bLm_{1}+\bh-2\blm_{2}+2\blm_{1}
-2\bt)_{i}
\ov2\mu_{i}})
Ch_{(\bh,\bt)}(\tld{q}).
\label{4.Bgtramp6}
\enr
Thus the spectrum is BPS due to supersymmetrized characters on the rhs.

 The other boundary states can be generated similar to (\ref{4.DBphi}).
It is easy to perform the calculation
similar to (\ref{4.Bgtramp6}) (see also ~\cite{ReS})
and find that tachyon decouples from the open string spectrum unless
(\ref{4.bps}).

 In conclusion of this section we note that the free-field 
representations of boundary states are 
determined modulo $BRST$-exact states satisfying $A$ or $B$-type boundary conditions.
We interpret this ambiguity in the free-field representation as a result of adding
brane-antibrane pairs annihilating under the tachyon condensation process
~\cite{Sen}. In this context the free-field representations of boundary states can be 
considered 
as the superpositions of branes flowing under the tachyon condensation to
nontrivial boundary states in Gepner models because they represents nontrivial $BRST$-homology
classes. It is also important to note that the automorphisms (\ref{1.reflect}) give
different free-field representations of boundary states because the corresponding
butterfly resolutions are not invariant with respect to these automorphisms.
However their cohomology are invariant. Hence these different representations have to be 
identified. Thus the free-field boundary states construction have to be rather
considered in derived category sense ~\cite{DCat}.

\vskip 10pt
\centerline{\bf 5. Open string spectrum and chiral de Rham complex.}

 In this section we represent an idea how the spectrum of open strings between the boundary states
can be investigated using free-field representations.
Our discussion will be very brief and restricted to the boundary states in
$\bmu=(3,3,3)$ model. For simplicity we shall ignore the space-time degrees of freedom
concentrating on the internal part of the spectrum.
We begin from the open string spectrum between $B$-type boundary states.

 The irreducible representations in $\mu=3$ minimal model can be labeled by $t=0,1,2$ so that
$M_{0}$ is vacuum representation with vanishing conformal weight and $J[0]$ charge, while
$M_{1}$ and $M_{2}$ are generated from $M_{0}$ by spectral flow operators $U^{-1}$ and $U^{-2}$.
Their highest vectors have conformal weights and $J[0]$ charges $(1/6,1/3)$ and $(1/6,-1/3)$
correspondingly. Thus in the product of minimal models characterizing by vector
$\bmu=(3,3,3)$ the set $\Dl'$ labeling the irreducible modules is
$\Dl'=\{(\bh,\bt)|h_{i}=0,\
t_{i}=0,...,2,\ i=1,...,3\}$.

 The set $\tld{\Dl}_{B}$ of solutions of (\ref{4.chargconj}) is given by
\ber
\tld{\Dl}_{B}=\{(\bh=(h,h,h),\bt=k\bv)|h=0,1, \ k=0,1,2\}
\label{5.conj}
\enr
The spectral flow invariant boundary states are numbered by the
spectral flow orbits $[\blm]=[\lm_{1},...,\lm_{3}]$, which are given by
\ber
[\blm]=\{[0,0,0],[1,2,0],[2,1,0]\}.
\label{5.Blms}
\enr
The $B$-type boundary states are given by (\ref{4.DB}). But it is easy to see that they
are equivalent to each other because $\sum_{i}\lm_{i}=0, mod3$.
Thus we have only one spectral flow invariant $B$-type boundary state.

 The transition amplitude is given by (\ref{4.Btramp9})
\ber
<<[\blm],\et,B|(-1)^{g}\exp{(\im 2\pi\tau(L[0]-{1\ov8}))}|[\blm],\et,B>>=
{1\ov3}\sum_{[\bt]}ch_{[\bt]}(-{1\ov\tau}),
\label{5.Bamp}
\enr
where
\ber
[\bt]=\{[0,0,0],[1,2,0],[2,1,0]\}.
\label{5.btorbits}
\enr

 We see that the total set of orbit characters appears in the amplitude.
Under the spectral flow $U^{{\bv\ov2}}$ the NS spectrum (\ref{5.Bamp}) goes to R spectrum which can
be interpreted geometrically.

 Note first that the spectrum of the open strings between spectral flow invariant boundary states 
is given by
$J[0]$ projection of the cohomology of the butterfly resolution
which is twisted by the spectral flow operators $U^{\bt+(n+{1\ov2})\bv}$.
The differential of the resolution is given by the sum
$\sum_{i}(Q^{+}_{i}+Q^{-}_{i})$ of screening charges.
Hence the calculation of cohomology can be given by two steps.

 At first step we can take for example
$\sum_{i}Q^{+}_{i}$ cohomology and then as the second step we calculate $\sum_{i}Q^{-}_{i}$
cohomology. It is well known (see for example ~\cite{FeS}, ~\cite{B})
that $\sum_{i}Q^{+}_{i}$ cohomology is generated by the set of $bc\bet\gm$-system of fields
\ber
a_{i}(z)=\exp{(\bs^{*}_{i}X(z))}, \
a^{*}_{i}(z)=(\bs_{i}\d X^{*}-\bs^{*}_{i}\psi\bs_{i}\psi^{*})
\exp{(-\bs^{*}_{i}X(z))}, \nmb
\al_{i}(z)=\bs^{*}_{i}\psi\exp{(\bs^{*}_{i}X(z))}, \
\al^{*}_{i}(z)=\bs_{i}\psi^{*}\exp{(-\bs^{*}_{i}X(z))}.
\label{5.btgm}
\enr
The spectral flow operators $U^{\bt}$ generate the twisted sectors for
these fields due to the relations
\ber
a_{i}(z_{1})U^{\bt}(z_{2})=
z_{12}^{-{t_{i}\ov\mu_{i}}}:a_{i}(z_{1})U^{\bt}(z_{2}):, \
a^{*}_{i}(z_{1})U^{\bt}(z_{2})=
z_{12}^{{t_{i}\ov\mu_{i}}}:a^{*}_{i}(z_{1})U^{\bt}(z_{2}):, \nmb
\al_{i}(z_{1})U^{\bt}(z_{2})=
z_{12}^{t_{i}-{t_{i}\ov\mu_{i}}}:\al_{i}(z_{1})U^{\bt}(z_{2}):, \
\al^{*}_{i}(z_{1})U^{\bt}(z_{2})=
z_{12}^{{t_{i}\ov\mu_{i}}-t_{i}}:\al^{*}_{i}(z_{1})U^{\bt}(z_{2}):.
\label{5.twist}
\enr
Let us consider in more details the space of states generated by the fields (\ref{5.btgm})
from the vacuum vector $|{1\ov2}\bv>$ which is generated by $U^{{1\ov2}\bv}$. In this sector
all the fields are expanded into integer modes and
\ber
a_{i}[n]|{1\ov2}\bv>=0, n>0, \
a^{*}_{i}[n]|{1\ov2}\bv>=0, n>-1,
\nmb
\al_{i}[n]|{1\ov2}\bv>=0, n>-1, \
\al^{*}_{i}[n]|{1\ov2}\bv>=0, n>0.
\label{5.annbtgm}
\enr
In terms of the fields (\ref{5.btgm}) the N=2 Virasoro superalgebra currents (\ref{2.Vird})
are given by
\ber
G^{-}=\sum_{i}\al_{i}a^{*}_{i}, \
G^{+}=\sum_{i}(1-{1\ov\mu_{i}})\al^{*}_{i}\d
a_{i}-{1\ov\mu_{i}}a_{i}\d\al^{*}_{i}, \nmb
J=\sum_{i}(1-{1\ov\mu_{i}})\al^{*}_{i}\al_{i}+{1\ov\mu_{i}}a_{i}a^{*}_{i},\nmb
T=\sum_{i}{1\ov2}((1+{1\ov\mu_{i}})\d\al^{*}_{i}\al_{i}-
(1-{1\ov\mu_{i}})\al^{*}_{i}\d\al_{i})+
(1-{1\ov2\mu_{i}})\d a_{i}a^{*}_{i}-
{1\ov2\mu_{i}}a_{i}\d a^{*}_{i}
\label{5.bgvir}
\enr
and it is easy to see that
\ber
J[n]|{1\ov2}\bv>=0, n>0, \ L[n]|{1\ov2}\bv>=0, n>0,
\nmb
G^{+}[n]|{1\ov2}\bv>=0, n>0, \
G^{-}[n]]|{1\ov2}\bv>=0, n>-1.
\label{5.annvir}
\enr
Hence $|{1\ov2}\bv>$ is the standard Ramound vacuum. Note that $G^{-}[0]$ is acting on the space of states
generated from this vector similar to de Rham differential action on the space of differential forms.
It is easy to see that the space of states generated by (\ref{5.btgm}) from $|{1\ov2}\bv>$ together with
$N=2$ Virasoro algebra action (\ref{5.bgvir})
has chiral de Rham complex structure introduced for any smooth manifold by 
Malikov, Schechtman and Vaintrob in ~\cite{MSV}. In our particular case
this manifold is $C^{3}$.
Adding the twisted sectors 
$U^{(n+{1\ov2})\bv}$ and making $J[0]$-projection converts the chiral de Rham complex
over $C^{3}$ into chiral de Rham complex on the orbifold ~\cite{GobM} of $C^{3}/Z_{3}$.
However the open string spectrum (\ref{5.Bamp}) contains also the
twisted sectors generated by $U^{(1,2,0)}$ and $U^{(2,1,0)}$ operators
which are not related to GSO projection. One can explain the appearance of these sectors as 
the result of
$B$-type boundary conditions. Indeed the charge conjugation condition (\ref{4.chargconj}) extracts the states
which are invariant modulo spectral flow shift. The charge conjugation $(\bh,\bt)\rightarrow (\bh,\bmu+\bh-\bt)$
does not commute with the GSO projection (simple current extension) giving thereby
the extension ~\cite{FSW}, ~\cite{GrP} of the orbifold group. 
Therefore $B$-type boundary states interact with the closed string states which are invariant
with respect to another copy of $Z_{3}$.
Thus they interact with the closed string states on $C^{3}/Z_{3}^{2}$ orbifold.

 Now we take the second step in the cohomology calculation.
The screening charges $Q^{-}_{i}$ can be expressed in terms of the fields (\ref{5.btgm}) as
\ber
Q^{-}_{i}=\oint dz \al_{i}a^{\mu_{i}-1}_{i}(z),
\label{5.LGscrn}
\enr
so $\sum_{i}Q^{-}_{i}$ is Koszul differential associated with Landau-Ginzburg potential 
$W(a_{i})=\sum_{i}a^{\mu_{i}}_{i}$
~\cite{FeS}, ~\cite{FrGLS}, ~\cite{Wit}.
Therefore $\sum_{i}Q^{-}_{i}$
cohomology calculation (the second step) corresponds to Landau-Ginzburg potential 
$\sum_{i}a_{i}(z)^{3}$ is 
switching on. Thus $B$-type boundary state is fractional brane of the
$C^{3}/Z_{3}^{2}$
Landau-Ginzburg orbifold.

 On the equal footing one can  consider first the $\sum_{i}Q^{-}_{i}$ cohomology.
It is given by "dual" $bc\bet\gm$-system of fields
\ber
b_{i}(z)=\exp{({1\ov\mu_{i}}\bs_{i}X^{*}(z))}, \
b^{*}_{i}(z)=(\mu_{i}\bs^{*}_{i}\d X-\bs_{i}\psi^{*}\bs^{*}_{i}\psi)
\exp{(-{1\ov\mu_{i}}\bs_{i}X^{*}(z))}, \nmb
\bet_{i}(z)={1\ov\mu_{i}}\bs_{i}\psi^{*}\exp{({1\ov\mu_{i}}\bs_{i}X^{*}(z))}, \
\bet^{*}_{i}(z)=\mu_{i}\bs^{*}_{i}\psi\exp{(-{1\ov\mu_{i}}\bs_{i}X^{*}(z))}.
\label{5.dubtgm}
\enr
The Virasoro currents are given by
\ber
G^{-}_{i}=(1-{1\ov\mu_{i}})\bet^{*}_{i}\d
b_{i}-{1\ov\mu_{i}}b_{i}\d\bet^{*}_{i}, \
G^{+}_{i}=\bet_{i}b^{*}_{i}, \nmb
J_{i}=-(1-{1\ov\mu_{i}})\bet^{*}_{i}\bet_{i}-{1\ov\mu_{i}}b_{i}b^{*}_{i},\nmb
T_{i}={1\ov2}((1+{1\ov\mu_{i}})\d\bet^{*}_{i}\bet_{i}-
(1-{1\ov\mu_{i}})\bet^{*}_{i}\d\bet_{i})+
(1-{1\ov2\mu_{i}})\d b_{i}b^{*}_{i}-
{1\ov2\mu_{i}}b_{i}\d b^{*}_{i}.
\label{5.dubgvir}
\enr

 It is easy to see that twisted sectors generated by the operators $U^{-\bt}$ for these fields coincide to the
twisted sectors for the fields (\ref{5.btgm}). It means in particular that the vector
$|-{1\ov2}\bv>$ satisfy the
same annihilation conditions (\ref{5.annbtgm}), with respect to the dual
fields (\ref{5.dubtgm}) but the fermionic part of (\ref{5.annvir}) is changed.
Thus it is the standard Ramound vacuum again.
Therefore the space of states generated by dual fields from $|-{1\ov2}\bv>$ together
with $N=2$ Virasoro algebra action (\ref{5.dubgvir}) is chiral de Rham complex on $C^{3}$.

 In the dual picture the $BRST$ operator $\sum_{i}Q^{+}_{i}$ plays the role of Koszul differential
\ber
Q^{+}_{i}=\oint dz \bet_{i}b^{\mu_{i}-1}_{i}(z).
\label{5.duLGscrn}
\enr
so we have the same Landau-Ginzburg potential in the dual coordinates.
Taking into account the $J[0]$ projection and twisted sectors one can conclude that we have
open string spectrum on the same Landau-Ginzburg orbifold.

 Note here that going from $bc\bet\gm$ fields (\ref{5.btgm}) to its dual
(\ref{5.dubtgm}) we change the role of fermionic $N=2$ Virasoro
superalgebra currents $G^{\pm}(z)$.
Initially $G^{-}[0]$ operator played the role of de Rham differential,
while in dual coordinates $G^{+}[0]$ becomes de Rham differential so that the chiral and anti-chiral rings
are exchanged also ~\cite{B}, ~\cite{B1}. 

 This "mirror" symmetry of the spectrum can be explained by toric geometry of the
chiral de Rham complex on the orbifold.
The set of screening charges $Q^{+}_{i}$ forming the differential at the first step
of cohomology calculations can be represented as the polytope in lattice $P\in \Pi$ whose vertices
are given by the origin $(0,0,0)$ and vectors $\bs_{i}$ from
$\Re$. The volume of the polytope measured in the units
of the lattice $P$ is $3^{3}$ and hence the group $Z_{3}^{3}$
acting in the product of minimal models can be represented as factor of $P$, by the sublattice
$P_{\Re}$ generated by the screening charge vectors $\Re$. The $J[0]$-projection is determined by the
vector $\bd-\bd^{*}\in \Pi$ and defines the sublattice
\ber
\Pi_{int}=\{(\bp,\bp^{*})\in \Pi|<\bp+\bp^{*},\bd-\bd^{*}>\in Z\}.
\label{5.intlat}
\enr
The intersection $P_{int}\equiv \Pi_{int}\cap P$ can be represented
\ber
P_{int}=\{\sum_{i}n_{i}\bw_{i}; n_{i}\in Z\}
\nmb
\bw_{1}=\bd,
\bw_{i+1}={1\ov 3}(\bs_{i+1}-\bs_{i}),\ i=1,2.
\label{5.intbas}
\enr
The intersection of the polytope $\Re$ with the height one plane
\ber
\Om=\{v\in E; <v,\bd^{*}>=1\}
\label{5.plan}
\enr
gives the following set of lattice points
\ber
\Sgm=\{\sgm_{0}=\bw_{1}, \sgm_{1}=\bw_{1}+\bw_{2}, \sgm_{2}=\bw_{1}+\bw_{3}, \sgm_{3}=\bw_{1}+\bw_{2}+\bw_{3},
\sgm_{4}=\bw_{1}-\bw_{2},
\nmb
\sgm_{5}=\bw_{1}-\bw_{3}, \sgm_{6}=\bw_{1}-\bw_{2}-\bw_{3},
\sgm_{7}=\bs_{1}, \sgm_{8}=\bs_{2}, \sgm_{9}=\bs_{3}\}
\label{5.trace}
\enr

 The lattice $P^{*}$ is dual to the lattice $P_{\Re}$.
The basic cone $K=\{\sum_{i}k_{i}\bs^{*}_{i}; k_{i}\geq 0\}\in P^{*}$ defines the set of monomials
$a^{k_{1}}_{1}...a^{k_{3}}_{3}$ on $C^{3}$ and hence the vertex algebra generated by the fields (\ref{5.btgm})
corresponds to the cone $K$ ~\cite{B}.
Taking into account $J[0]$-projection one can see ~\cite{As}
that toric manifold determined by mutually dual lattices $P_{int}$, $Hom(P_{int},Z)$
and the polytope $\Re$ is singular with the orbifold
singularity $Z_{3}^{2}$. It allows to identify the divisors $\sgm_{0},...,\sgm_{6}$
with the operators $U^{\bt}$ arriving thereby "(twist-field)-divisor" map
~\cite{As}-~\cite{AsGM}. For example the operators $U^{\bv}$, $U^{(1,2,0)}$, $U^{(2,1,0)}$
correspond
to the divisors $\sgm_{0},\sgm_{5}, \sgm_{6}$.

 According to the construction of Borisov ~\cite{B} one can resolve the orbifold singularity
~\cite{As} and obtain chiral de Rham complex on the smooth CY manifold ~\cite{B}
adding to the differential of the butterfly resolution the screening charges
$\oint dz \sgm_{0}\psi^{*}\exp(\sgm_{0}X^{*})(z),...,
\oint dz \sgm_{6}\psi^{*}\exp(\sgm_{6}X^{*})(z)$ associated to the divisors
$\sgm_{0},...,\sgm_{6}$. 

 In the "dual" picture we associate the polytope $\Re^{*}$ in the lattice $P^{*}$
to the set of screening charges $Q^{-}_{i}$. Let us denote by $P^{*}_{\Re^{*}}$ the sublattice
in $P^{*}$ generated by $\Re^{*}$. The factor $P^{*}/P^{*}_{\Re^{*}}$ gives the group $Z_{3}^{3}$
of symmetries of the minimal models. In the lattice $P^{*}_{int}\equiv P^{*}\cap \Pi_{int}$
we fix the basis $\bw^{*}_{i}$
\ber
\bw^{*}_{1}=\bd^{*}, \ \bw^{*}_{i+1}=\bs^{*}_{i+1}-\bs^{*}_{i}, \
i=1,2.
\label{5.intbas1}
\enr

 The intersection of the polytope $\Re^{*}$ with the height one plane
\ber
\Om^{*}=\{v^{*}\in E^{*}; <v^{*},\bd>=1\}
\label{5.duplan}
\enr
gives the following set of lattice points from $P^{*}_{int}$
\ber
\Sgm^{*}=\{\sgm^{*}_{0}=\bw^{*}_{1}, \sgm^{*}_{1}=\bw^{*}_{1}+\bw^{*}_{2},
\sgm_{2}=\bw^{*}_{1}+\bw^{*}_{3},
\sgm^{*}_{3}=\bw^{*}_{1}+\bw^{*}_{2}+\bw^{*}_{3},
\sgm^{*}_{4}=\bw^{*}_{1}-\bw^{*}_{2},
\nmb
\sgm^{*}_{5}=\bw^{*}_{1}-\bw^{*}_{3},
\sgm_{6}=\bw^{*}_{1}-\bw^{*}_{2}-\bw^{*}_{3},
\sgm^{*}_{7}=3\bs^{*}_{1}, \sgm^{*}_{8}=3\bs^{*}_{2}, \sgm^{*}_{9}=3\bs^{*}_{3}\}
\label{5.dutrace}
\enr

 The lattice $P$ is dual to the lattice $P^{*}_{\Re^{*}}$.
The cone $K^{*}=\{\sum_{i}k_{i}\bs_{i}; k_{i}\geq 0\}\in P$ defines the set of monomials
$b^{k_{1}}_{1}...b^{k_{3}}_{3}$ on $C^{3}$ and hence the vertex algebra generated by the fields (\ref{5.dubtgm})
corresponds to the cone $K^{*}$.
Taking into account $J[0]$-projection we can see that toric manifold determined by the lattices
$P^{*}_{int}$, $Hom(P^{*}_{int}, Z)$ and the polytope
$\Re^{*}$ is singular with the orbifold
singularity $Z_{3}^{2}$. One can also associate the divisors $\sgm^{*}_{0},...\sgm^{*}_{6}$
to the operators $U^{-\bt}$. For example the operators
$U^{-\bv}$, $U^{-(1,2,0)}$, $U^{-(2,1,0)}$ correspond
to the divisors $\sgm^{*}_{0},\sgm^{*}_{5}, \sgm^{*}_{6}$.
Thus the "mirror" background for the open string is the same.
The resolution of this manifold and construction of the corresponding chiral de Rham complex on the
smooth CY manifold are given similar. This time one has to add to the differential of
the butterfly resolution the screening charges
$\oint dz \sgm^{*}_{0}\psi\exp(\sgm^{*}_{0}X)(z),...,
\oint dz \sgm^{*}_{6}\psi\exp(\sgm^{*}_{6}X)(z)$.

 Now we consider the open string spectrum between $A$-type boundary states. As is well known they
correspond in the large volume limit to the special Lagrangian submanifolds ~\cite{BeBeS} which are real
submanifolds in CY manifold. In the free-field representation this fact manifests in the
relations (\ref{3.Aantz}), (\ref{3.AX}) between
the left-moving and right-moving degrees of freedom for $A$-type boundary conditions.

 $A$-type boundary states in $\bmu=(3,3,3)$ model which are spectral flow invariant 
are labeled by
 (\ref{5.Blms}) and the set $\Dl_{CY}$ is given by
 \ber
[\bt]=\{[0,0,0],[1,2,0],[2,1,0]\}.
\label{5.orbits}
\enr

Let us consider first the transition amplitude $Z^{A}_{[\blm],[\blm]}$, for spectral flow
invariant boundary state. From (\ref{4.amp7}) we obtain
\ber
Z^{A}_{[\blm],[\blm]}(\tau)=
\nmb
<<[\blm],\et,A|(-1)^{g}\exp{(\im 2\pi\tau(L[0]-{1\ov8}))}|[\blm],\et,A>>=
ch_{[0,0,0]}(-{1\ov\tau}).
\label{5.Aamp}
\enr

 Analogously to the $B$-type case the spectrum (\ref{5.Aamp}) is given by the cohomology
of the corresponding butterfly resolution which is calculated by two steps.
At first step we see that in R sector $\sum_{i}Q^{+}_{i}$-cohomology are given by
$bc\bet\gm$ system of fields (\ref{5.btgm}) and the space of states generated by these fields
from the Ramound vacuum $|{1\ov2}\bv>$ has the chiral de Rham complex structure on $C^{3}$.
Adding the twisted sectors $U^{(n+{1\ov2})\bv}$ and making $J[0]$ projection converts the chiral de Rham complex
over $C^{3}$ into chiral de Rham complex over the orbifold $C^{3}/Z_{3}$.
At the second step of the calculation one has to take the cohomology with respect to
the differential $\sum_{i}Q^{+}_{i}$ which is the Koszul differential associated 
to the potential $W=\sum_{i}a^{3}_{i}$.

 The transition amplitude between different spectral flow invariant boundary states
 gives additional twisted sectors.
Let us take, for example $[0,0,0,]$ and $[1,2,0]$. Then we obtain
\ber
<<[0,0,0],\et,A|(-1)^{g}\exp{(\im 2\pi\tau(L[0]-{1\ov8}))}|[1,2,0],\et,A>>=
ch_{[1,2,0]}(-{1\ov\tau}).
\label{5.amp1}
\enr
Before we take $\sum_{i}Q^{-}_{i}$-cohomology the spectrum of states is generated by the fields
(\ref{5.btgm})
from the twisted vacuum vectors $|(n+{1\ov2})\bv+\bt>$, where $\bt=(1,2,0)$.
These additional twisted states correspond to the open string ending
on different D-branes in $C^{3}/Z_{3}$ orbifold such that twisting determines the angle.
The angles can be read off easily from (\ref{5.twist})
\ber
a_{1}(e^{\im 2\pi}z)=a_{1}(z), \
a_{2}(e^{\im 2\pi}z)=e^{\im{2\pi\ov3}}a_{2}(z), \
a_{3}(e^{\im 2\pi}z)=e^{-\im{2\pi\ov3}}a_{3}(z), \nmb
a^{*}_{1}(e^{\im 2\pi}z)=a^{*}_{1}(z), \
a^{*}_{2}(e^{\im 2\pi}z)=e^{-\im{2\pi\ov3}}a^{*}_{2}(z), \
a^{*}_{3}(e^{\im 2\pi}z)=e^{\im{2\pi\ov3}}a^{*}_{3}(z), \nmb
\al_{1}(e^{\im 2\pi}z)=\al_{1}(z), \
\al_{2}(e^{\im 2\pi}z)=e^{\im{2\pi\ov3}}\al_{2}(z), \
\al_{3}(e^{\im 2\pi}z)=e^{-\im{2\pi\ov3}}\al_{3}(z), \nmb
\al^{*}_{1}(e^{\im 2\pi}z)=\al^{*}_{1}(z), \
\al^{*}_{2}(e^{\im 2\pi}z)=e^{-\im{2\pi\ov3}}\al^{*}_{2}(z), \
\al^{*}_{3}(e^{\im 2\pi}z)=e^{\im{2\pi\ov3}}\al^{*}_{3}(z).
\label{5.monodr}
\enr

 The similar analysis can be carried out in terms of the "dual" $bc\bet\gm$-system of fields.
It gives the equivalent picture of D-branes on $C^{3}/Z_{3}$ Landau-Ginzburg
orbifold if we take into account "mirror" involution of the $N=2$ Virasoro superalgebra.

 In this example one can see the general property of $A$-type branes that
the operator
$\exp{(-\im2\pi\sum_{i}\phi_{i}J_{i}[0])}$ rotates the $D$-branes on the orbifold
and $\blm$ parameterize the angles. One can check this by the direct calculation

 Let us find the boundary condition for the spectral flow
operator $U^{\bt}$.
It belongs to (left-moving) CY Hilbert space if it has integer $J[0]$ charge.
We have
\ber
U^{\bt}|[\bLm,\blm],\et,A>>=
\exp{(-\im2\pi\sum_{i}\lm_{i}J_{i}[0])}
\exp{(\im2\pi\sum_{i}{2t_{i}\lm_{i}\ov\mu_{i}})}U^{\bt}|[\bLm,0],\et,A>>=
\nmb
\exp{(-\im2\pi\sum_{i}{t_{i}(\Lm_{i}-2\lm_{i})\ov\mu_{i}})}\bar{U}^{-\bt}|[\bLm,\blm],\et,A>>,
\label{5.thetang}
\enr
where it is implied that
$J[0]$ charge of the state $U^{\bt}$ is integer. We have
used here the quasi-isomorphism property of spectral flow
operator. In other words we identified to each other the Ishibashi states
which differ by the action of spectral flow operator
$\prod_{i=1}^{r}U_{i}^{n_{i}\mu_{i}}\bar{U}_{i}^{n_{i}\mu_{i}}$, where
$n_{i}$ are integers . Taking into account these identifications we find that
$A$-type boundary state is localized at the angles
$\tt_{i}=-2\pi{[\bLm-2\blm]_{i}\ov\mu_{i}}$.

 In conclusion of this section we would like to note first
that the discussion carried out in this section can be generalized to the case of
boundary states in general Gepner models. Thus, the orbifold structure of the open
string spectrum appears in general case which allows to interpret directly the Gepner models
boundary states as the fractional branes on the Landau-Ginzburg orbifolds.

 The second remark is related to the intersection index and $D$-branes charges calculation in the
boundary state approach ~\cite{DFi}, ~\cite{BDLR}-~\cite{NaNo}. This calculation in boundary state
formalism is given by two steps. At first step we calculate the closed string transition amplitude
which is given by the linear combination of characters. At the second step one has to take a
limit of these expression when $Im\tau$ goes to infinity. As a result only chiral primary fields
in the open string sector give a contribution to the intersection index. As we have argued in this
section the open string spectrum has the chiral
de Rham complex structure. It was shown by Malikov Schechtman and Vaintrob ~\cite{MSV}
that chiral de Rham complex is a (loop-coherent) sheaf of vertex algebras ~\cite{B}.
Then one can give geometric interpretation of the first step as a
calculation of the index of the Dirac operator ~\cite{EOTY}, ~\cite{Ell}-~\cite{Ell2},
~\cite{EllB} of the sheaf.
So it is natural to suggest that it gives thereby string generalization of the bilinear
form on the $K$-theory classes.

\vskip 10pt
\centerline{\bf Acknowledgments}

 I would like to thank Jan Troost for his interest to this work and helpful discussions.
I am very grateful to LPTHE of University Paris 6 and CPHT of Ecole Polytechnic
for the hospitality where the last part of this work was done.

 This work was supported in part by grants RBRF-04-02-16027,
RBRF-2044.2003.2, INTAS-OPEN-03-51-3350.

\end{document}